\documentclass[amsfonts,aps,pre,twocolumn]{revtex4}
\usepackage{graphicx}

\usepackage{array}
\usepackage{amsmath}

\input{macros.inp}
\input{abbk.inp}
\renewcommand{\tauphi}{\eww{\tau}}
\renewcommand{\tauphiT}{\eww{\tau(T)}}

\newcommand{\gammaia}{\gamma_{i,\alpha}}
\newcommand{\Eia}{E_{i,\alpha}}
\newcommand{\Eik}{E_{i,\alpha(k)}}
\newcommand{\Eikk}{E_{i(k),\alpha(k)}}
\newcommand{\pik}{p_{i,\alpha(k)}}
\newcommand{\Eappest}{E_\ind{app}^\ind{est}}
\newcommand{\Efst}{\bar E_{01}}
\newcommand{\pia}{p_{i,\alpha}}
\newcommand{\pbb}{p_\ind{BB}}
\newcommand{\Eappi}{E_{\ind{app}}(i)}
\renewcommand{\equ}[1]{Eq.~\ref{#1}}

\begin{document}
\title{
Energy Barriers and Activated Dynamics in a Supercooled Lennard-Jones Liquid
}
\author{B. Doliwa$^1$ and
A. Heuer$^2$}
\affiliation{$^1$Max Planck Institute for Polymer Research, 55128 Mainz, Germany}
\affiliation{$^2$Institute of Physical Chemistry, University of M\"unster -
  48149 M\"unster, Germany}
\date{\today}
\begin{abstract}
We study the relation of the potential energy landscape (PEL) topography
to relaxation dynamics of a small model glass former of Lennard-Jones type.
The mechanism under investigation is the hopping betweem superstructures
of PEL mimima, called metabasins (MB).
From the mean durations $\tauphi$ of visits to MBs,
we derive effective depths
of these objects by the relation $\Eapp=\d\ln\tauphi/\d\beta$, where $\beta=1/\kB T$.
Since the apparent activation energies $\Eapp$ are of purely dynamical origin,
we look for a quantitative relation to PEL structure.
A consequence of the rugged nature of MBs
is that escapes from MBs
are not single hops between PEL minima, but complicated multi-minima sequences.
We introduce the concept of return probabilities to the bottom of MBs in order
to judge whether the attraction range of a MB was left. We then compute the
energy barriers that were surmounted.
These turn out to be in good agreement with the effective depths $\Eapp$, calculated from
dynamics.
Barriers are identified with the help of a new method, which accurately performs
a descent along the ridge between two minima.
A comparison to another method is given.
We analyze the population of transition regions between minima, called basin borders.
No indication for the mechanism of diffusion to change around the mode-coupling transition
can be found.
We discuss the question whether the one-dimensional reaction paths connecting
two minima are relevant for the calculation of reaction rates at the temperatures under study.
\end{abstract}
\maketitle


\section{Introduction.}

More than thirty years ago, Goldstein~\cite{Goldstein:1969}
proposed to view a glass-forming system as a point moving in the
high-dimensional landscape of the potential energy $V(x)$.  In
this framework he suggested to focus onto the local minima of the
potential energy landscape (PEL), where the system is supposed to be trapped at
low enough temperatures.  Via occasional transitions to
neighboring minima the system finally relaxes. Owing to the
separation of time scales, one is able to describe many
features of glass formers
by properties of only the minima.
Stillinger and Weber~\cite{Stillinger:222}
formulated this idea in the language of statistical thermodynamics using the concept
of basins. A basin of a given minimum is defined as the set of
configurations that reach this minimum via their steepest descent
path $\dot{x}=F(x)$.  (We set $x$ and $F(x)$ as shorthands for the
multidimensional vectors all particle positions and all forces, respectively.)
The resulting tiling of configuration space into
different basins allows one to write the free energy approximately
as a function of static properties of minima, i.e. their energies
and vibrational frequencies~\cite{Buchner:193,Sciortino:59}.
Knowledge of the thermodynamics is in
general not sufficient to predict dynamical properties like
diffusion constants or relaxation times. However, experimental~\cite{Richert:218}
as well as simulated~\cite{Saika:247,Sastry:198,Scala:71}
data seem to indicate that there exists a strong connection
between dynamics and thermodynamics via the Adam-Gibbs relation~\cite{Adam:39}.

Our goal is to reach a quantitative understanding of the slowing down of dynamics,
as expressed
by the bulk long-time diffusion constant $D(T)$. Mode coupling
theory (MCT)~\cite{Gotze:1992} predicts a power-law behavior of the
form $1/D(T)\propto(T-T_c)^\gamma$ above the MCT critical
temperature $T_c$. Since $T_c$ is found to be higher than the glass
transition temperature $T_g$, the MCT divergence of $1/D(T)$ at
$T_c$ is not observed in practice. The common explanation for this
shortcoming of MCT is that the theory neglects `activated
processes', or 'hopping', which is supposed to come into play around
and below $T_c$. Indeed it was proven by Schr\o der
et~al.~\cite{Schroder:88} that in the vicinity of $T_c$ the time
scale of fast local dynamics around single minima becomes well
separated from the time scale of interbasin transitions. Above
$T_c$ the common picture suggests that the dynamics are
'entropy-driven', i.e. characterized by the search for escape
directions~\cite{Donati:211,Scala:375,LaNave:265}, since saddles lie far below
the instantaneous potential energy of the system and thus represent
no serious barriers~\cite{Angelani:212,Angelani:315}.
Also different observables like the average order of
saddles~\cite{Angelani:212,Broderix:228} and the number of
diffusion-like normal modes ~\cite{Donati:211,Scala:375,LaNave:265,Scala:376}
seem to indicate that well above $T_c$ the dynamics is not
governed by activated transitions between adjacent minima.
However, in
the multi-dimensional space of particle coordinates, it is not
obvious how to distinguish thermally activated from entropically
limited dynamics. One possible way to do this will be discussed
below.
For the time being,
we use the term hopping only in a formal sense,
meaning that the trajectory of the system is mapped
onto a sequence of jumps among minima.

Looking for a quantitative link of bulk diffusion $D(T)$ to
PEL properties, we recently investigated hopping dynamics
on the PEL in greater detail~\cite{Doliwa:341}.
A priori, temporal and spatial aspects of hopping events
have to be considered, the former in the shape of the
waiting time distribution (WTD) of jumps, the latter by the jump
lengths and directions, and correlations thereof.
We found that
strong backward correlations of jumps arise from the organization
of minima into superstructures, which,
following~\cite{Stillinger:333}, we call metabasins (MB). It had
already been known from previous work~\cite{Buchner:11} that
structural relaxation corresponds to jumps among MBs rather than
single basins. MBs were identified with the help of a
straightforward algorithm such that close-by minima between which
the system performs several back- and forth jumps are identified as
a single MB.
Then, indeed, hopping among MBs was found to be close to a random
walk with a distribution
of MB waiting times.
Motivated by this fact, we expressed $D(T)$ in the simple form,
\NEQU{D(T)=\frac{a^2}{6N\tauphiT},}{EQUDIFFTAU}
with the mean waiting time $\tauphiT$ and the {\it effective}
jump length $a(T)$.
With this ansatz, we anticipated that waiting times would carry the major
part of the temperature dependence.
Indeed, $a(T)$ turned out to be constant for $T<2T_c$,
which is why we dropped the argument of $a(T)$.
\equ{EQUDIFFTAU} constitutes an important step towards the understanding
of diffusion in supercooled liquids:
it suffices to look for the physics behind MB waiting times,
spatial details of hopping being expressed by a single constant.

A simple model for hopping dynamics has been discussed
by Bouchaud and coworkers~\cite{Monthus:310}.
They consider the relaxation from traps of depths $E$ with distribution $\rho(E)$,
and escape rates $\gamma(E,T)=\gamma_0\exp(-\beta E)$.
When
$\rho(E)\propto\exp(-E/T_x)$, the WTDs assume
power-law tails $\psi(\tau)\propto\tau^{-\alpha(T)}$ with exponents
$\alpha(T)=1+T/T_x$. The consequence is the divergence of the mean
waiting time at temperature $T_x$.
In our recent paper, we observed that the WTDs
of a binary Lennard-Jones system
are in conformance with this kind of
power-law decay~\cite{Doliwa:341}.
As a consequence of such slowly decaying WTDs, the mean value
$\tauphiT$ was found to be dominated by the few, very long waiting
times. In other words, the temperature dependence of $D(T)$ follows
alone from the durations of trapping in the very stable MBs. These
results were obtained for small binary Lennard-Jones mixtures of
$N=65$ particles. For a macroscopic system, which, due to its
dynamic heterogeneity~\cite{Sillescu:60}, contains many slow and
fast subsystems in parallel, this implies the dominance of slow
regions in the temperature dependence of $D(T)$.

The logical continuation along this line of thinking is to relate
MB lifetimes to the PEL topography.
The most prominent
characteristics of a MB is, of course, its energy $\epsmb$,
which is defined as the lowest energy of all its constituent
minima. It is then natural to introduce the mean MB lifetime
$\tauepsmbT$ at constant $\epsmb$.
Knowledge of $\tauepsmbT$, together with the population of MBs, $p(\epsmb;T)$,
is sufficient to calculate $\eww{\tau(T)}$ and thus $D(T)$, as we will show now.
We write
\NEQU{\tauphiT=\int\d\epsmb\tauepsmbT\phi(\epsmb;T),}{EQUDIFFTAUEPS}
where $\phi(\epsmb;T)$ is the distribution of MBs visited at
temperature $T$. We will see that this
decomposition can be achieved by a detailed analysis of the
hopping dynamics.
Since $p(\epsmb;T)$ denotes the probability that at a
given time the system is in a MB with energy $\epsmb$, it is
proportional to $\phi(\epsmb;T)$ and the time $\tauepsmbT$ the
system remains in MBs of this energy. With the appropriate
normalisation one gets
\NEQU{p(\epsmb;T)=\frac{\tauepsmbT}{\tauphiT}\phi(\epsmb;T).}{EQUPPHI}
From \equthree{EQUDIFFTAU}{EQUDIFFTAUEPS}{EQUPPHI},
it immediately follows the representation
\NEQU{D(T)=\frac{a^2}{6N}\eww{\frac{1}{\tauepsmbT}}_T.}{EQUDIFFTDPEL}
Here, $\eww{...}_T$ denotes the canonical time average (w.r.t. $p(\epsmb;T)$), while
$\eww{...}$ is the average over MBs.
Hence,
\NEQU{\brac{\tauepsmbT,\ p(\epsmb;T)}\rightarrow\eww{\tau(T)}\rightarrow D(T),}{EQUCONCEPT}
where the second implication has been established in our recent paper~\cite{Doliwa:341}.
The population $p(\epsmb;T)$ is related to the single-basin population $p(\eps;T)$,
a purely static quantity,
which has been extensively discussed in the literature~\cite{Sciortino:59,Sastry:198,Buchner:193}.
It has turned out for Lennard-Jones mixtures that the number density $\Geff(\eps)$ of minima
is approximately gaussian.
Thus, the population of minima, $p(\eps;T)\propto\Geff(\eps)e^{-\beta\eps}$, could be expressed
by three parameters describing global PEL structure.
In the present paper, we focus on $\tauepsmbT$, our goal being
to deduce it from PEL structure.
If this succeeds, we have
established the following connection,
\EQU{\tm{local + global PEL structure}\rightarrow\tm{long-time dynamics},}{}
which, in our opinion, pushes the understanding of diffusion in supercooled liquids a step further.

We proceed as follows. We first compute MB lifetimes from ordinary simulation,
and later compare them to the prediction from PEL structure.
First, we characterize the relaxation from four single, randomly selected MBs.
By an exhaustive sampling of these MBs, we will be able to get some first insights
into MB topology.
Second, many MBs of fixed energy are considered and their lifetimes $\tauepsmbT$
are calculated.
Third, we relate MB lifetimes to PEL structure, by quantifying
the MB depths, or effective barriers, which determine
the temperature dependence of $\tauepsmbT$.
The physical scenario which will emerge from the results of this
paper implies that MBs can be regarded as traps, surrounded by
high barriers. It turned out from exhaustive explorations of PEL
connectivity~\cite{Doye:216} that due to the high dimensionality
of configuration space the number of escape paths from every
minimum is enormous. Thus, one may anticipate that the effective
barrier to leave a specific MB results as a complex superposition
of individual escape paths. Therefore, enormous numerical effort is
required to quantify their multitude for many different MBs.

Note that the whole analysis will be carried out in the spirit
of activated barrier crossing.
The extent to which this is present in supercooled liquids is quite disputed in literature.
However, we will show that for temperatures
in the landscape-influenced regime
below $2T_c$, the apparent activation energy
\NEQU{\Eapp(\epsmb;T)=\frac{\d}{\d\beta}\ln\tauepsmbT,}{EQUEPSEAPP}
can indeed be identified with PEL barriers much
larger than $k_B T$ which the system encounters when leaving a MB.
Thus, together with \equ{EQUDIFFTDPEL}, we will find that the activated escape
out of deep traps is the physical mechanism behind diffusion.

To our knowledge, such a connection between dynamics and PEL barriers
has never been established for a fragile glass former.
In contrast, for $\tm{SiO}_2$, the apparent activation energy of
diffusion below $T_c$ could be related to the simple breakage of $\tm{Si}-\tm{O}$
bonds~\cite{Mcmillan,Horbach}.

The organization of the paper is as follows. In
section~\ref{SECSIM}, we provide the details of our simulation, and
describe the interval bisection method to identify MBs.
Section~\ref{SECMTAU} deals with the computation of apparent
activation energies from relaxation dynamics. The corresponding
energy barriers will be addressed in section~\ref{SECBARRIERS},
after introducing our technique for finding transition states
(section~\ref{SECNLRM}). In section~\ref{SECACT}, we
independently demonstrate
that barriers and associated reaction paths indeed govern
relaxation. Finally, we discussion further aspects of our results
in section~\ref{SECDISCUSSION} and conclude in
section~\ref{SECCONCLUSION}.


\section{Simulation Details.}
\label{SECSIM}
\subsection{General.}
In the present work, we investigate a binary mixture of
Lennard-Jones particles (BMLJ), as recently
treated by two groups~\cite{Broderix:228,Wales:286};
see also~\cite{Kob:203}.
It is characterized by the interaction
potentials
\EQU{V_{\alpha\beta}(r)=4\epsilon_{\alpha\beta}[(\sigma_{\alpha\beta}/r)^{12}-(\sigma_{\alpha\beta}/r)^{6}]}{}
with the parameter set $N=N_A+N_B=52+13=65$,
$\sigma_{AB}=0.8\sigma_{AA}$, $\sigma_{BB}=0.88\sigma_{AA}$,
$\epsilon_{AB}=1.5\epsilon_{AA}$, $\epsilon_{BB}=0.5\epsilon_{AA}$, $r_c=1.8$.
Linear
functions were added to the potentials to ensure continuous forces
and energies at the cutoff $r_c$.
These modifications of the original potential by Kob and Andersen~\cite{Kob:203}
are necessary for the simulation of small systems.
We use Langevin molecular
dynamics simulations (MD) with fixed step size,
$\lambda^2=0.015^2=2\kB T\Delta t/m\zeta$,
equal particle masses $m$,
friction constant $\zeta$ set to unity,
and periodic boundary conditions.
Units of length, mass, energy, and time are $\sigma_{AA}$, $m$, $\epsilon_{AA}$,
and $m\zeta\lambda^2/2\epsilon_{AA}$, respectively.
However, we will omit these units, for convenience.
The mode-coupling temperature is
$T_c=0.45\pm0.01$ in this model system (compare~\cite{Kob:203}).
For the analysis of
dynamics from the PEL perspective it is essential to use small
systems, as has been stressed in the literature~\cite{Buchner:11,Keyes:335,Grigera:264}.
On the other hand, naturally, the system
should not be too small in order to avoid major finite-size effects.
For the BMLJ, $N \approx 60$ turns out to be a very good
compromise~\cite{Broderix:228,Wales:286,Buchner:11},
whereas $N\le 40$ already causes large finite-size effects
~\cite{Buchner:193}.
Here we choose $N=65$, since the BMLJ60 system
has a stronger tendency to be trapped in crystalline configurations.
We stress here that the results obtained for the BMLJ65 system
show no finite-size related artifacts.
For example, $D(T)$ of the BMLJ1000 is identical to $D(T)$ of the BMLJ65
above $T_c$, see \figref{FIGMTAU}.
In the temperature range studied, we found that the behavior
of a BMLJ130 system largely resembles
that of two independent copies of a BMLJ65.
Thus, the generalization of the present work to larger systems should
not bear any pitfalls.

Interestingly, the BMLJ65 relaxation becomes Arrhenius-like for
low temperatures.
Since we have no equilibrium runs of our BMLJ1000 below $T_c$
it is unclear whether the Arrhenius behavior of the BMLJ65 down to $T=0.4$
is a finite-size effect.
A possible explanation is that
the lower end of the PEL is reached (located at $\epsmin\approx-302$, see \figref{FIGEPS}),
preventing further increase of barriers.
In turn, this may be related to the fact that cooperative regions cannot grow any further
in the BMLJ65, i.e. structural optimization 
- which happens upon cooling - finally comes to an end.
Since we work above $T_c$, our key results are not affected by this argument.

\subsection{Interval bisection.}
By regularly quenching the MD trajectory $x(t)$ to the bottom of the
basins visited at time $t$, as proposed by Stillinger and Weber,
we obtain a discontinous trajectory $\xmin(t)$.
A problem from the standpoint of simulations is to resolve the
{\it elementary} hopping events.  Since computer time prohibits to
calculate the minimum $\xmin(t)$ for every time step $t$, we
normally find ourselves in the situation of having equidistant
quenched configurations $\xmin(t_i)$, $t_i=i\Delta t$, with, say,
$\Delta t=10^5$~MD~steps. If the same minimum is found for times
$t_i$ and $t_{j}$, we need not care about transitions in the
meantime, because no relaxation has occurred there. If, in
contrast, $\xmin(t_i)\ne\xmin(t_{i+1})$, we must not expect
$\xmin(t_{i+1})$ to be the direct successor of $\xmin(t_{i})$,
since many other minima could have been visited between $t_i$ and
$t_{i+1}$. Therefore, further minimizations in this time interval
are necessary. For reasons of efficiency, we apply a
straightforward interval bisection method, which locates
transitions to an accuracy of 1~MD~step: provided
$\xmin(t^{(0)}_\ind{start})\ne \xmin(t^{(1)}_\ind{start})$,
(a)~set $t^{(0)}\lar t^{(0)}_\ind{start}$, $t^{(1)}\lar
t^{(1)}_\ind{start}$, (b)~reconstruct the trajectory $x(t)$ at
time
$t^{(2)}=(t^{(0)}+t^{(1)})/2$, (c)~calculate
$\xmin(t^{(2)})$, (d)~if $\xmin(t^{(2)})=\xmin(t^{(0)})$, set
$t^{(0)}\lar t^{(2)}$, else set $t^{(1)}\lar t^{(2)}$, (e)~repeat
(b)-(d) until $t^{(1)}-t^{(0)}=1$~MD~step. Repeated application of
the interval bisection to a simulation run $x(t)$ finally gives
all relevant transitions. Note that the determination of {\it all}
transitions including the numerous recrossings of basin borders
would require minimization for {\it every MD~step}! The interval
bisection method thus may oversee back- and forth motions between
minima which, in any event, are irrelevant for relaxation.
Although computationally demanding, the above method has proved
most efficient for resolving the relevant details of hopping on the PEL
and is predestined for the construction of metabasins (see below).


\section{Activation Energies from Metabasin Lifetimes.}
\label{SECMTAU}

\subsection{Metabasin lifetime construction.}

As said above, stable configurations in the supercooled liquid
are rarely due to single minima on the PEL, but mostly correspond
to groups of strongly correlated minima. While the system is
trapped in such a MB for a long time, a small number of minima is
visited over and over again. This is well reflected by the time
series of potential energies,
$\eps(t)=V(\xmin(t))$~\cite{Buchner:11,Doliwa:341}. In this
section, we will dwell on the computation of mean MB lifetimes, (i)
for single, selected MBs, and (ii) averaged over MBs of a given
energy $\epsmb$, thus yielding $\tauepsmbT$.
The individual MBs of~(i)
correspond to long-lived MBs and thus represent typical MBs which
govern the temperature dependence of $\tauphiT$.

For the grouping of minima into MBs and the resulting determination
of their lifetimes from a regular simulation run, we use the following,
straightforward algorithm ~\cite{Buchner:11}.
(a) determine the regions
$[t_i^*,t_i^\dagger]$
where $t_i^*$ is the time of
the first and $t_i^\dagger$ the time of the last occurrence of
minimum $\xmin(t_i)$,
(b) any two regions overlapping by less than
$\taumol$ are cut so that
$[t_i^*,t_i^\dagger]\cap[t_j^*,t_j^\dagger]=\emptyset$, where
$\taumol$ is a small molecular time-scale,
(c) any two regions
overlapping by more than $\taumol$ are combined to
$[t_i^*,t_i^\dagger]\cup[t_j^*,t_j^\dagger]$,
(d) the lifetimes of MBs are defined by the regions after step (c),
(e) the MB energy $\epsmb$ is defined as the lowest energy
of minima visited during the MB lifetime.

A few comments on the procedure are in order.  Time regions in (a)
are determined by the interval bisection method which yields the
time of transition from one minimum to another with an accuracy of
one MD~step. Step (b) is motivated by the observation that
recrossings of a basin border during a transition are very
probable. If we ignore this fact, i.e. set $\taumol=0$, step (c)
would merge nearly all regions and we would end up with
unphysically long MBs. Instead, we use $\taumol=40$,
which is the basin equilibration time obtained from energy autocorrelation.
Since the durations of transitions are of the same order of magnitude as $\taumol$,
this is a sensible choice.
However, we found that the results for MB lifetimes are not very susceptible
to the precise value of $\taumol$.
Step (c) itself is the realization
of the MB concept. It is important to note that, different
from~\cite{Buchner:11}, we will treat all MBs on the same footing
here, no matter if they are  short-lived or long-lived.

So far, the MB lifetime construction rests upon single trajectories, which
only partially reflect the configuration space topology.
In section~\ref{SECBARRIERS}, the MB concept
will be given a more precise, static definition, based on the return probability
to the ground minimum.

\subsection{Activation Energies for Single MBs.}
\label{SECMTAUFUNNEL}
\figany{\WiePosit}{\includegraphics[width=8cm]{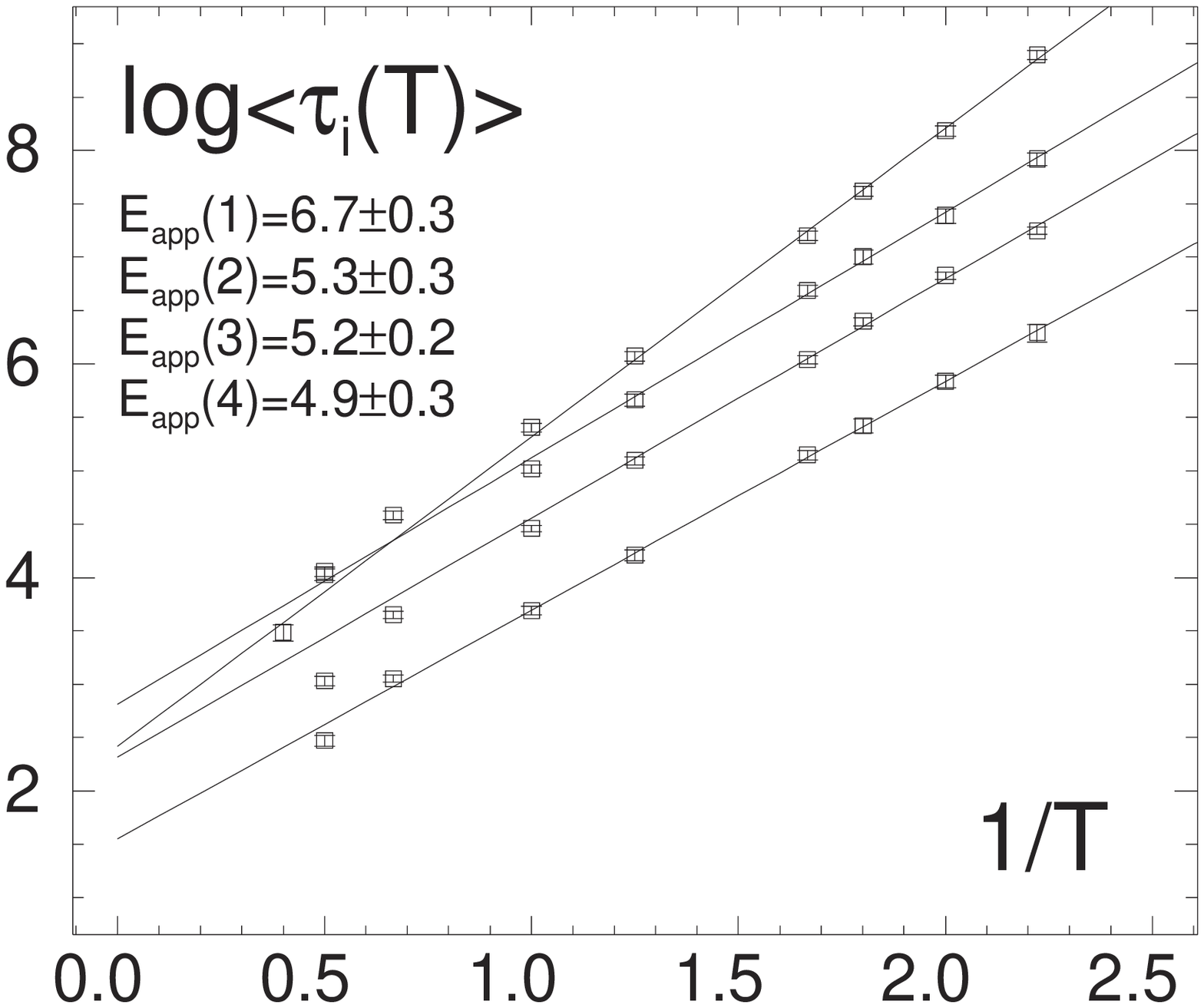}}{
Mean lifetimes of four low-lying, randomly selected metabasins,
computed from repeated escape runs
($\epsmb=-301.64, -300.47, -300.16,$ and $-300.74$, from top to bottom).
The number of runs are 85,59,175, and 105, from top to bottom.
Arrhenius fits work well in the temperature range $T\le1\approx2.2T_c$,
the corresponding activation energies are given in the figure.
Curves have been shifted vertically by $0.5(4-i)$ orders of magnitude
for better inspection.
}{FIGTAUFUNNEL}

As noted above, the temperature dependence of $D(T)$ is dominated
by the long-lived MBs. Generally, these are low-lying MBs, i.e. deep traps in the PEL.
Since different MBs differ in their stability, a statistical treatment will be needed.
As a first step, however,
we restrict ourselves to the investigation of single MBs.

The relaxation times computed in this section do not stem from
regular, linear simulation runs, but are obtained by artifically
placing the system in  a specific MB and waiting for its
escape ('escape runs'). The above algorithm for the MB lifetime
construction implicitly assumes that MBs finally have been left.
In other words, the algorithm may not be used to determine the
time where to stop the simulation due to successful escape.
Fortunately, we can avoid running into this paradoxical situation
by judging from an independent criterion whether an escape has been completed:
if the distance of the instantaneous minimum to the starting position
is greater than $\dmax=4$,
returning to the original basin can practically be excluded
(see section~\ref{SECBARRIERS} for a justification of $\dmax=4$).
Then, by applying the MB construction algorithm to the escape run,
we obtain the lifetime of the MB.

We analyzed four low-lying ($\epsmb<-300$), randomly selected MBs in greater detail.
By repeated starts from the bottom of the MBs,
we computed the mean lifetimes $\eww{\tau_i(T)}$ as a function of temperature.
From~\figref{FIGTAUFUNNEL}, we see that the relaxations
from all MBs follow nicely an Arrhenius law.
We note that, due to starting at minima, a short
intra-basin equilibration time ($\taumol=40$, from energy autocorrelation)
has been subtracted from the raw $\eww{\tau_i(T)}$.

The fact that an Arrhenius form of $\tauepsmbT$ is observed
indicates that the barriers do not change any further upon
lowering temperature.
Put differently, MBs serve as
traps surrounded by barriers with heights around
$\Eappi=\d\ln\eww{\tau_i(T)}/\d\beta$.  We will see in
section~\ref{SECBARRIERS} that this is indeed correct. Since
$\Eappi/\kB T_c>10$, this implies a strongly activated dynamics
near $T_c$.

\subsection{Activation Energies \vs MB energies.}
As a further step, we analyze the mean relaxation time from MBs
with the same energy, $\tauepsmbT$; see \equ{EQUDIFFTAUEPS}.
Clearly, the low $\epsmb$ are not populated at high temperatures
so that regular simulation does not yield $\tauepsmbT$ over a wide
temperature range. We therefore artificially place the system in
the desired MBs (in the lowest minima $\epsmb$ thereof) and
measure the escape times as a function of temperature. Averaging
over many different MBs, we obtain $\tauepsmbT$.
\figany{t}{\includegraphics[width=7cm]{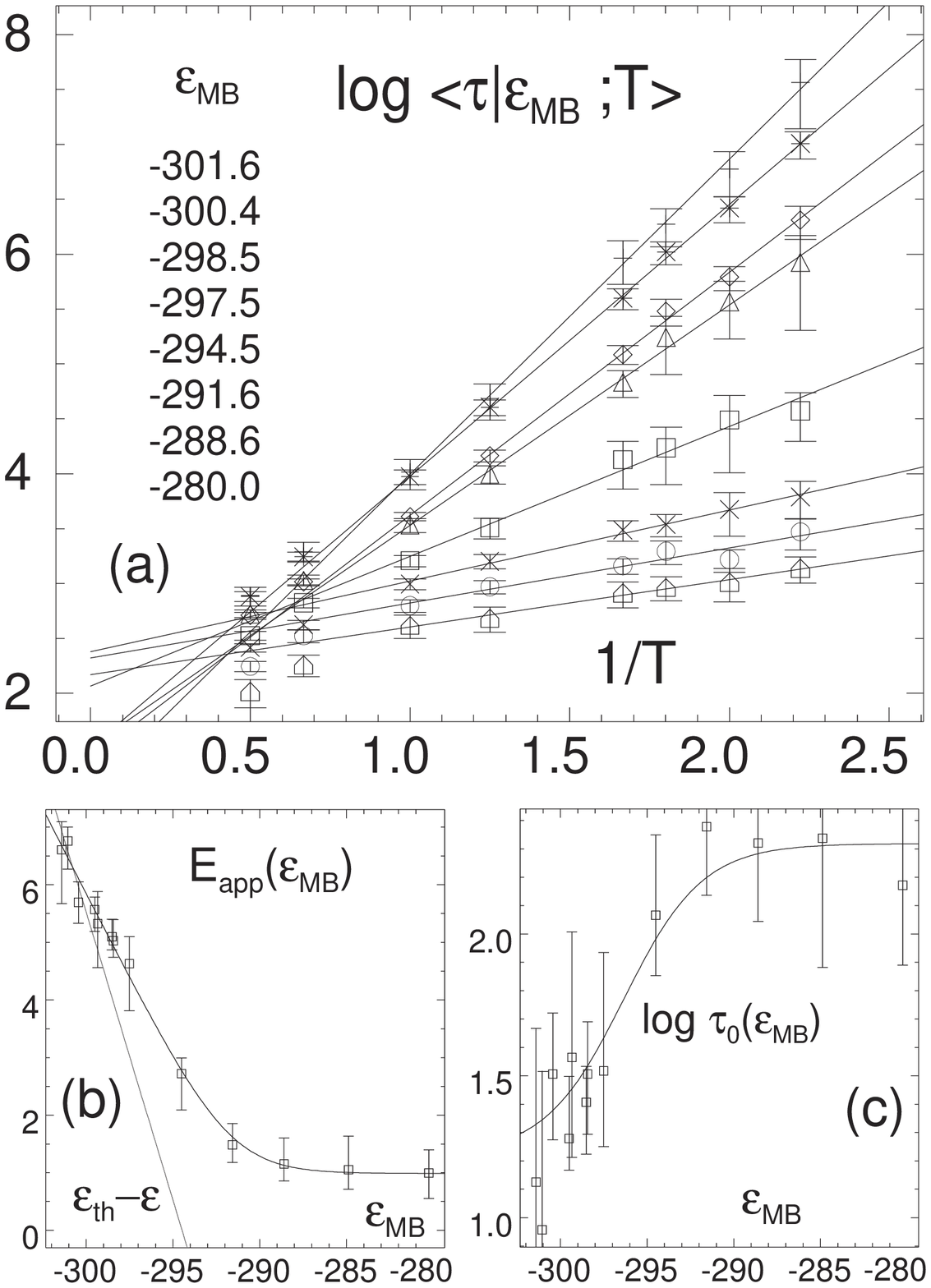}}{ (a) Arrhenius
plot of mean MB lifetimes $\tauepsmbT$, for different $\epsmb$. A
MB equilibration time of $\taumol=40$ was subtracted. Straight
lines are fits of the form \equ{EQUTAUEPSMB}.
(b) apparent activation
energies $\Eapp(\epsmb)$.
(c)
prefactors $\tau_0(\epsmb)$. Curved lines are interpolations of
the data. }{FIGTAU}
Results are shown in \figref{FIGTAU} as a function of $\epsmb$.
Below $T=1$, all relaxation times display Arrhenius behavior. Thus the
apparent activation energies $\Eapp(\epsmb;T)$ are temperature
independent. In the following we will therefore omit the
second argument. Thus, we can write
\NEQU{\tauepsmbT=\tau_0(\epsmb)e^{\beta\Eapp(\epsmb)}.}{EQUTAUEPSMB}
As expected, the properties of MBs as expressed by
$\Eapp(\epsmb)$ depend on their ground state energy $\epsmb$.

We can interpret $\Eapp(\epsmb)$ as the mean effective depth of
MBs at $\epsmb$. Since the lower end of the energy landscape is
reached at $\eps\approx-302$
no deeper traps exist
(compare \figref{FIGEPS}, see also~\footnote{B.~Doliwa and A.~Heuer, in preparation}).
A simple statement for the depths of traps
would follow if the rims of all traps were at the same level
$\epsth$.
The consequence
would be $\Eapp(\epsmb)=\epsth-\epsmb$, for all $\epsmb<\epsth$.
This simple scenario is ruled out by the data, see
\figref{FIGTAU}(b).
Actually, a more complicated energy dependence of $\Eapp(\epsmb)$
is expected from the very fact that the system - despite its small size -
is not a completely cooperative unit, see the discussion in section~\ref{SECDISCUSSION}.

The fact that we still observe Arrhenius-like relaxation in
\figref{FIGTAU} indicates that the variation of trap depths at
constant $\epsmb$ is not large, compare $\Eappi$ from
\figref{FIGTAUFUNNEL}. Otherwise, $\Eapp(\epsmb)$ would increase
upon decreasing temperature, due to the more and more dominant,
extremely deep traps. In contrast, trap depths at constant
$\epsmb$ seem rather well defined by $\epsmb$, which suggests the
existence of some underlying topological principle.

As seen from \figref{FIGTAU}(c), the prefactor $\tau_0(\epsmb)$
has no strong dependence on $\epsmb$.
From high energies, it decreases at most an order of magnitude and seems
to level off below $\epsmb=-297$.
Hence, for the range of energies that dominate $\tauphiT$ at low temperatures,
it can be considered constant within error bars.
In contrast to $\Eapp(\epsmb)$, we will not be able to deduce $\tau_0(\epsmb)$
from PEL structure. Its weak variation is therefore quite fortunate.

\figany{t}{\includegraphics[width=7.5cm]{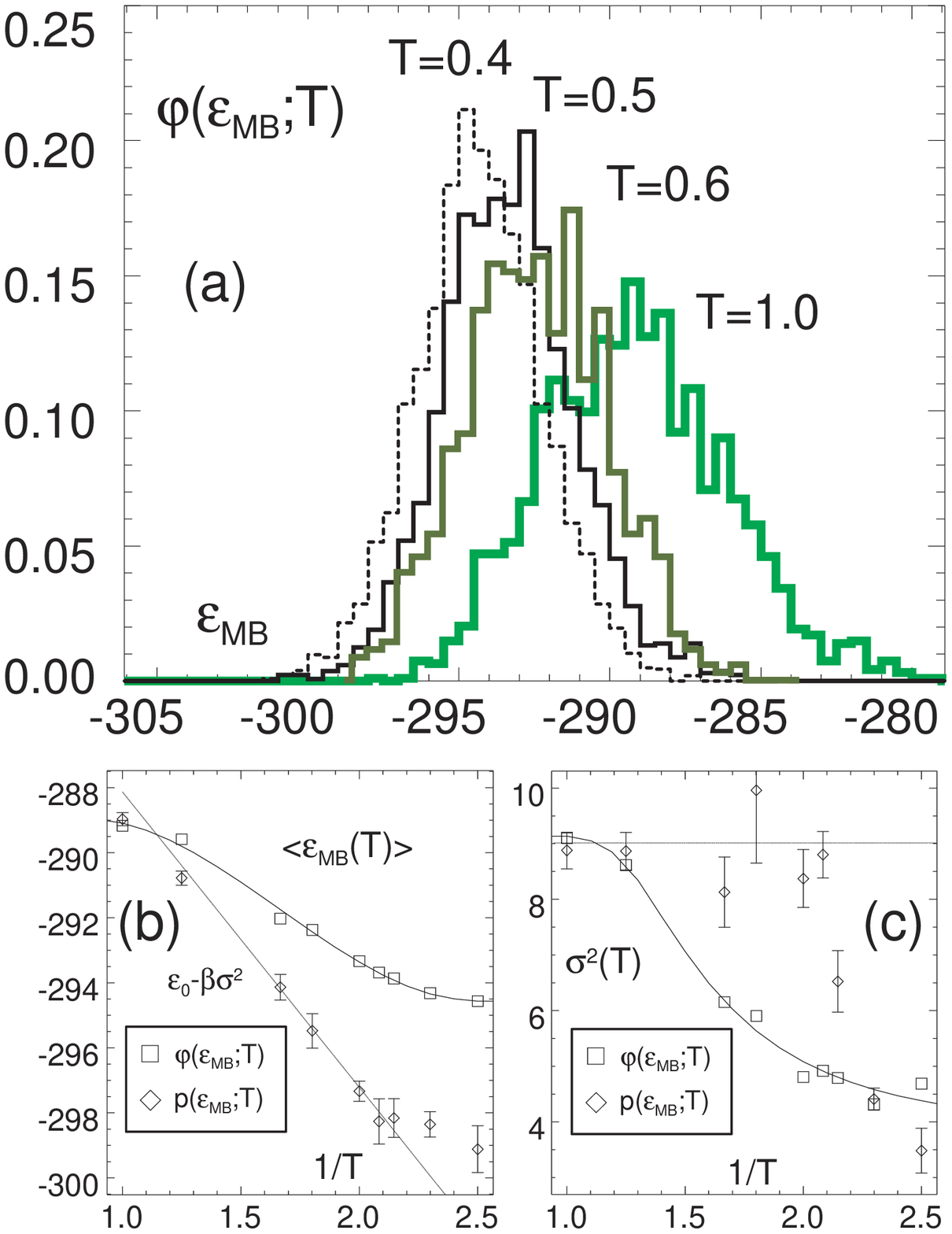}}{
(a) distribution $\phi(\epsmb;T)$ of MB energies, for four temperatures.
(b) Mean energies, from $\phi(\epsmb;T)$ and from $p(\epsmb;T)$.
(c) Variances of the distributions $\phi$ and $p$.
Polynomial fits to the data are shown in (b) and (c).
Straight lines are predictions for $p$ from an ideally gaussian distribution $\Geff(\epsmb)$
(mean $\eps_0$, variance $\sigma^2$).
Deviations are due to reaching the lower end of the PEL,
i.e. the deepest amorphous minima.
}{FIGEPS}
We will now analyze the second factor of the integrand in
\equ{EQUDIFFTAUEPS}, $\phi(\epsmb;T)$. It is shown in
\figref{FIGEPS}. Interestingly, the variation of $\phi(\epsmb;T)$
is much weaker for low $T$ as the variation of $p(\epsmb;T)$.
From \equtwo{EQUPPHI}{EQUTAUEPSMB}, one concludes that the
constancy of the distribution $\phi(\epsmb;T)$ is equivalent
to having $\Eapp(\epsmb) = \epsth - \epsmb$, with some constant $\epsth$.
Since this simple behavior is not present, one must still have a
residual temperature dependence of $\phi(\epsmb;T)$.

Concerning $p(\epsmb;T)$,
the weak
temperature dependence of its first moment for the
three lowest temperatures is simply related to the probing of the
lower end of the PEL.
Actually, it turns out that  $p(\epsmb;T)$ is, within
statistical error, identical to the corresponding distribution of
minima $p(\eps;T)$.
One would expect this for high $\epsmb$, because no pronounced MBs
are observed there.
Considering a deep MB with many minima, this will equally effect no
large difference between $p(\epsmb;T)$ and $p(\eps;T)$.
The reason is that the 
group of minima near $\epsmb$ carry the largest part of the population.
Since they are close to $\epsmb$, transferring their weight
to $\epsmb$ when computing $p(\epsmb;T)$ has little effect.

\figany{\WiePosit}{\includegraphics[width=8cm]{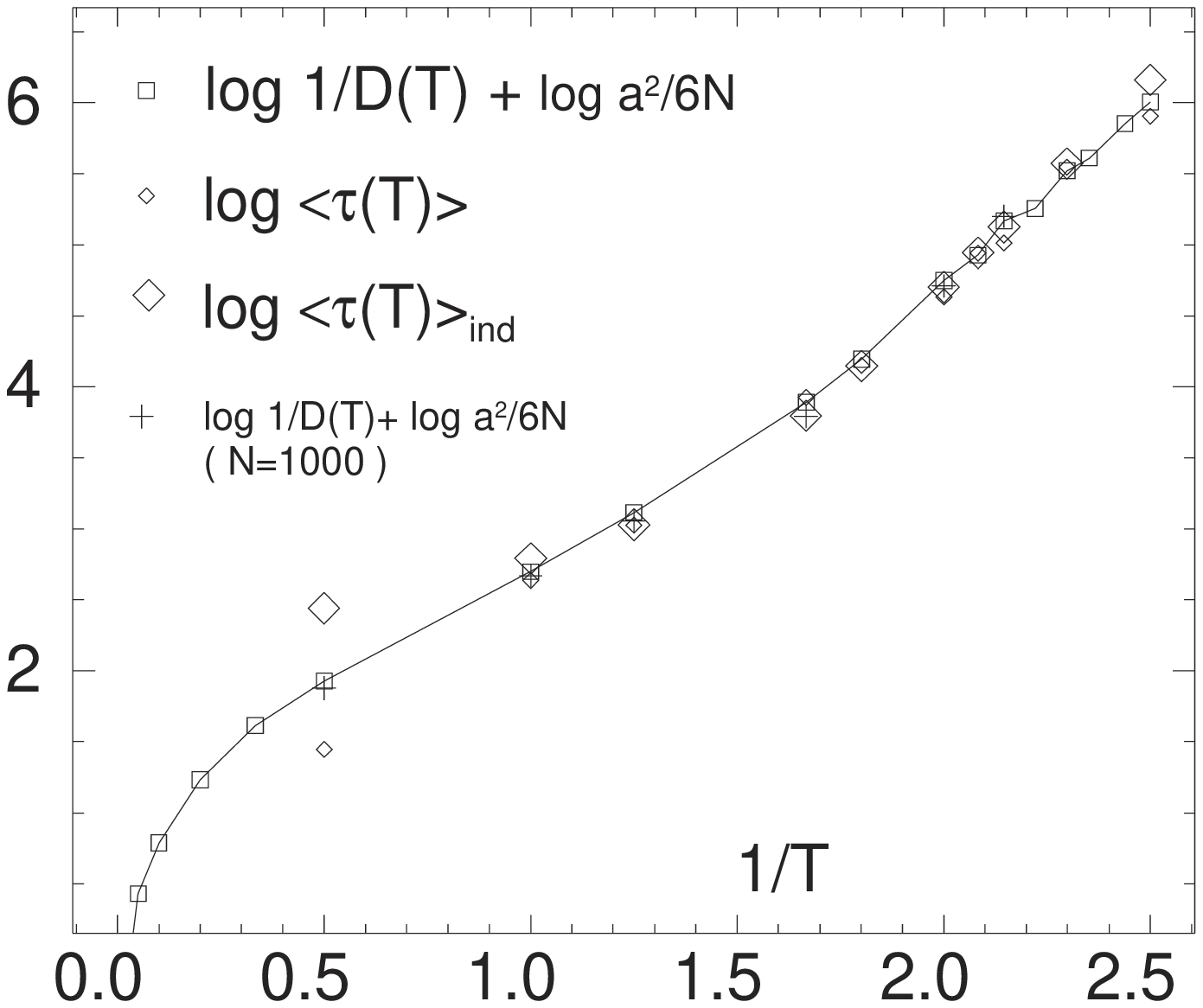}}{
Arrhenius plot of the mean waiting time $\tauphiT$
versus the indirectly determined counterpart,
$\tauphiTind$.
For comparison, we also show
the inverse one-particle diffusion
constant $1/D(T)$ multiplied by a constant ($a^2=1.0$), see~\cite{Doliwa:341}.
Error bars are of the order of the symbol size.
Also included is the $1/D(T)$ for the BMLJ1000 system.
}{FIGMTAU}
As a consistency check, we use the data from \figref{FIGTAU} and \ref{FIGEPS}
to reproduce $\tauphiT$ indirectly via \equ{EQUDIFFTAUEPS}
(denoted $\tauphiTind$).
The match
with $\tauphiT$ is not completely trivial since the data for
$\tauphiT$ and $\phi(\epsmb;T)$ were gathered from a linear simulation run,
while $\tauepsmbT$ results from selected MBs of certain $\epsmb$,
where the system has been artificially placed.
As shown in \figref{FIGMTAU}, the agreement of $\tauphiT$ and $\tauphiTind$
is good for $T\le1$ within the possible accuracy.
Note that there is no free fit parameter between them.
The deviation at $T=2$ can be explained by the fact that $\tauepsmbT$,
above $T=1$, and especially for the high $\epsmb$, departs from Arrhenius behavior,
see \figref{FIGTAU}(a).

So far, all barriers or trap depths have been derived indirectly,
from the temperature dependence
of waiting times.
A link to the PEL structure is still lacking.
For instance, the activation energies $\Eapp(i)$ of this section
are expected to reflect the local topography of the selected MBs.
Indeed, they can be identified from the barriers of escape paths,
as will be demonstrated in section~\ref{SECBARRIERS}.

First of all, the barriers between neighboring minima are of great
interest. These are known once we have in hand the corresponding
transition states.


\section{Non-Local Ridge Method for finding Transition States.}
\label{SECNLRM}

\subsection{Description of the method.}

We now describe how to determine transition states (TS) from
the simulation, by what we call
the (non-local) ridge method. The principle idea is that TSs are
local minima of basin borders. They can be pictured as the lowest
points of mountain ridges on the PEL. If the system crosses a
basin border at time $t$, the steepest descent path starting from
$x(t)$ should end up in a TS, see~\cite{Ionova:294}. In practice,
however, the descent will deviate from the ridge due  to
numerical error, finally ending up in the minimum
$\xi_0\equiv\xi(t-)$ or $\xi_1\equiv\xi(t+)$. As a way out, we let
the system perform two descents in parallel, on either side of the
basin border, as schematically depicted in \figref{FIGSCHEMATS}.
More specifically, if a transition happened at time $t$, interval
bisection yields the configurations $x(t)\equiv y_0$ and
$x(t+1\tm{ MD step})\equiv y_1$. From these, by further interval
bisection on the straight line between $y_0$ and $y_1$, the
distance to the border may be further reduced if necessary,
resulting in two configurations, again called $y_0$ and $y_1$.
Close as they are, they still belong to different basins.
If we now let descend $y_0$ and $y_1$ in parallel, they first move
along the ridge towards the transition state until they finally
bend off to their respective minima. This separation is clearly
not wanted, so from time to time we reduce their distance by
interval bisection.
After a few iterations (descents+interval bisection)
the vicinity of the transition state is
reached in most cases. We then use a short minimization of the
auxiliary potential $\tilde V=\frac12|F(x)|^2$ followed by a few
steps of Newton-Raphson type, which bring the search for the TS to
a quick convergence. Besides a vanishing force, the resulting
configuration $\zeta$ has a Hessian matrix with one negative
eigenvalue. After small displacements along the corresponding
eigenvector, one reaches the adjacent minima via steepest descent.
This yields
the reaction path (RP) $\zeta(s)$,
where $s$ is a curvilinear parameter.
We set $\zeta(0)=\zeta$, $\zeta(s_0)=\xi_0$, and $\zeta(s_1)=\xi_1$,
where either $s_0$ or $s_1$ is negative.

\figany{\WiePosit}{\includegraphics[width=6cm]{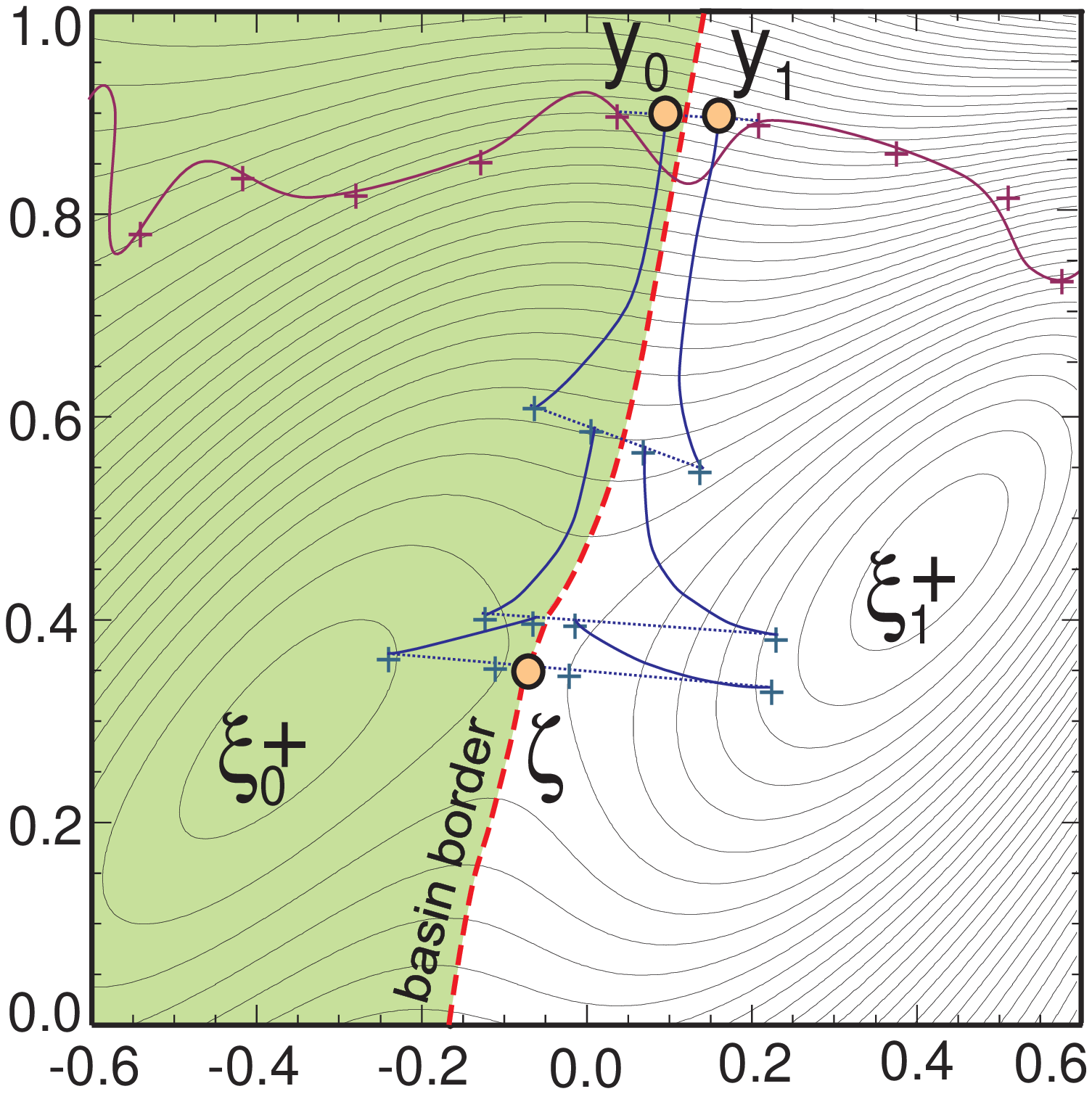}}{
Sketch of the TS search with the
ridge method.
}{FIGSCHEMATS}
It can happen, though, that no saddle between $y_0$ and $y_1$ is
found, but that the interval bisection locates a third minimum.
The basin border splits into two at this point, and no direct
saddle between the initial and final minimum is available. Thus,
we also have to split the descent along the basin border into two
processes and then continue separately. If the two descents are
successful without further bifurcations, we are finished and have
the optimum reaction path which takes a detour via a third
minimum. In such a situation, the RP is clearly not very useful.
It has to be stressed that bifurcations are no artifacts of the
ridge method, but a topological feature of some basin borders on
the PEL. Fortunately, as a signature of strong anharmonicity, they
are quite rare and happen to occur only in the high-energetic
regions of the PEL. For the escapes from long-lived MBs, they are
of no importance.

A similar algorithm is described in the literature~\cite{Ionova:294},
which, instead of minimization and interval bisection, uses local
maximization between $y_0$ and $y_1$ to prevent the configurations from moving apart.
Although computationally less expensive, this method is not appropriate for our purpose.
As an effect of the high dimensionality, the local shape of the PEL
around $y_0$ and $y_1$ gives no direct clue to the membership to basins.
When descending, one may thus loose the important property
of $y_0$ belonging to the basin of $\xi_0$ and $y_1$ belonging to that of $\xi_1$.
This effect has indeed been reported in~\cite{Ionova:294}.

In the literature, plenty of methods exist dealing with the
computation of transition states. One kind of them starts from the
knowledge of the initial and final
minimum~\cite{Elber,Matro,Deaven,Angelani:13}. After a more or
less educated guess for an initial trial RP, one iteratively
improves the RP according to some prescription, e.g., the
minimization of an action functional. Two sources of erroneous
results have to be addressed in this connection. First, the two
minima in question have to be true neighbours. This can only be
verified by locating two points close to the basin border, e.g. by
interval bisection of the initial trial path. The numerical cost
is not small; for our ridge method, for instance, about one third
of the calculation time is consumed by fixing $y_0$ and $y_1$
(depending on the minimization interval of the MD run). Second,
the iterative path optimization may become stuck in a local
extremum, due to an unfortunate choice of the initial path.

The other kind of TS search methods start from an initial minimum
and climb up to a transition state guided by the shape of the PEL.
Just walking against the force, however, would be a fatal strategy, as
one can see by turning the PEL upside down:
ending up in a TS is numerically impossible, since
one quickly runs into one of the PEL singularities
(two or more identical particle positions). Eigenvector-following
algorithms~\cite{Wales:239} overcome this defocussing of steepest
ascent paths by walking into the direction of negative local PEL
curvature. The 'activation-relaxation technique' by Mousseau and
coworkers, in contrast, steps against the force in the direction
leading away from the minimum, while descending the PEL
perpendicular to that direction~\cite{Barkema:131}. A drawback of
the latter methods is that the choice for the next TS to mount is
not well under control. From the minimum, a starting direction is
chosen, either by purely random displacements or by some
hard-sphere-like particle moves~\cite{Doye:216}. Unfortunately,
the number of escape directions from a minimum is generally very
large (at least $O(Nd)$ as we found in the BMLJ65), whereas the
majority of these is dynamically inaccessible at low $T$. Hence,
eigenvector-following and activation-relaxation techniques yield
many TSs which only negligibly contribute to relaxation rates.
Striving for the simulation of low-temperature hopping dynamics
based on these methods~\cite{Wales:286,Mousseau:44,Ball:382}, one
may suffer a considerable reduction of efficiency. 

We finally mention two complementary means of studying energy barriers.
The 'lid' algorithm, proposed by Sch\"on and coworkers~\cite{Wevers},
is able to find upper bounds for the depths of single basins.
By performing random walks below different potential energy thresholds
and by regular minimizations, one is able to compute the elevation necessary
for transitions to neighboring minima.
From a more theoretical perspective, Schulz has specified
a relation between transition rates and
the overlap of vibrations in neighbouring basins~\cite{Schulz:321}.

\subsection{Comparison to $\tilde V$-saddles.}
\label{SECITALO}
The advantage of the ridge method is that we definitely find the
relevant barrier for a transition, i.e. a first order saddle on
the basin border next to the point where the border was crossed.
In contrast, the method using the auxiliary potential $\tilde
V=\frac 12 |F(x)|^2$ as applied in recent
studies~\cite{Angelani:212,Angelani:315,Broderix:228,Grigera:264}
has two major drawbacks.
First, the $\tilde V$ minimization locates saddles
(we call them $\tilde V$-saddles),
even if they are not accessible kinetically.
This is because the
expression $F^\dagger HF$ is not positive ($H=H(x)$
denotes the Hessian of $V(x)$),
i.e. $\tilde V$-minimization can
{\it climb up} to a saddle.
Second, one obtains higher-order saddles
and, most frequently, non-stationary points (shoulders).
These configurations are of no use to us
because we specifically analyze paths over the lowest barriers
on basin borders, i.e., transition states.

\figany{\WiePosit}{\includegraphics[width=6cm]{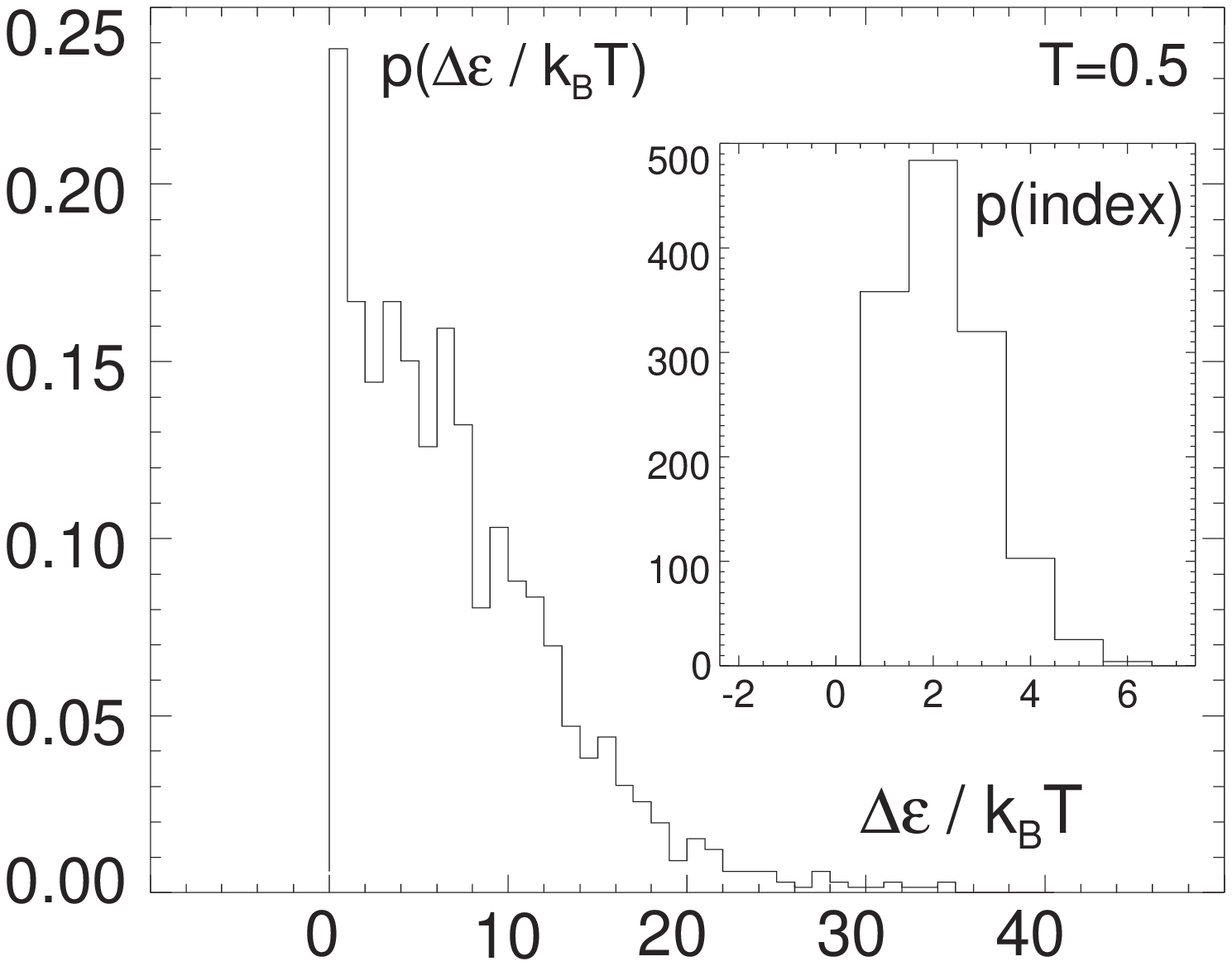}}{
Comparison of transition states, obtained via the ridge method,
with minima of the auxiliary potential
$\tilde V$. Starting points for saddle computations lay close to basin borders.
Main plot: histogram of $\tilde V$-saddle minus TS energies.
Inset: histogram of indices of $\tilde V$-saddles.
}{FIGITALOVSBURKI}
To shed more light on the interrelation of TSs and $\tilde V$-minima,
we minimized $\tilde V$ by steepest descent,
starting from
configurations $x(t)$ only if $\xi(t)\ne\xi(t+1\tm{ MD step})$
(like $y_0$ in \figref{FIGSCHEMATS}).
In other words, we calculated $\tilde V$-saddles exactly at transition times.
If this yielded the correct TSs, our more time-consuming
ridge method would be clearly useless.
The difference $\Delta\eps=\epssad-\epstst$ specifies the overestimation
of the true barrier by the $\tilde V$-saddle.
It may also happen that the index of the $\tilde V$-saddle is different from one.
The distributions of $\Delta\eps$ and the index are shown in
\figref{FIGITALOVSBURKI} ($T=0.5$).
Obviously, the $\tilde V$-saddles considerably overestimate
barriers and the correct TSs are
only found very rarely.
Moreover, most of the $\tilde V$-saddles have an index different from one, i.e. are no TSs at all.
In turn, the energy of the TS is never undersold by a $\tilde V$-saddle.
In conclusion, $\tilde V$-saddles turn out to have the undesired quality of
being decorrelated from the relevant TSs, i.e.,
from the barriers that control relaxation (see section~\ref{SECACT}).

\subsection{Population of Basin Borders.}
After Angelani and coworkers~\cite{Angelani:212,Angelani:315},
the mean index of $\tilde V$-saddles vanishes at $T_c$.
Therefore, as they have argued, dynamics above $T_c$ is dominated by saddles, in that
there are always some unstable directions available which allow the system to relax,
without traversing an additional energy barrier.
Passing $T_c$, the mechanism suffers a drastic change, and abruptly, one is faced with an index
of ca. zero, i.e., saddles have to be reached via thermal activation.
Since the preceeding subsection may cast some doubts on the significance of $\tilde V$-saddles,
we now want to discuss an alternative analysis of the way the population of minima
versus unstable configurations
evolves upon decreasing temperature. More specifically, we determine the population of basin borders,
\NEQU{\pbb(T) = \frac{1}{Z(T)}\int\d\mathcal B\int\d xe^{-\beta V(x)}\delta(x-\mathcal B),}{EQUPBB}
where integration is over the non-crystalline part of configuration space,
also in the partition function $Z(T)$,
and $\mathcal B$ runs over all basin borders of the PEL.
This expression is impractical in numerical simulation; one may rather ask if, for some
instantaneous configuration $x$, there is a basin border nearby.
In this case, small random displacements
(length $\delta\in\reell$, direction $\omega\in\reell^{Nd}$, $|\omega|=1$) possibly
lead into another basin, i.e. $\xi(x)\ne\xi(x+\omega\delta)$.
This kind of PEL analysis has been recently carried out
by Fabricius and Stariolo~\cite{Fabricius:381}.
One calculates
\NEQU{\pbb(T;\delta)=\eww{P\bra{\xi(x)\ne\xi(x+\omega\delta)}}_{T,\omega},}{EQUPBB2}
which is the probability that random disturbances $\omega\delta$ will
cause crossings of basin borders at temperature $T$. The brackets denote the canonical
plus the average over the random directions $\omega$.
\figany{\WiePosit}{\includegraphics[width=8cm]{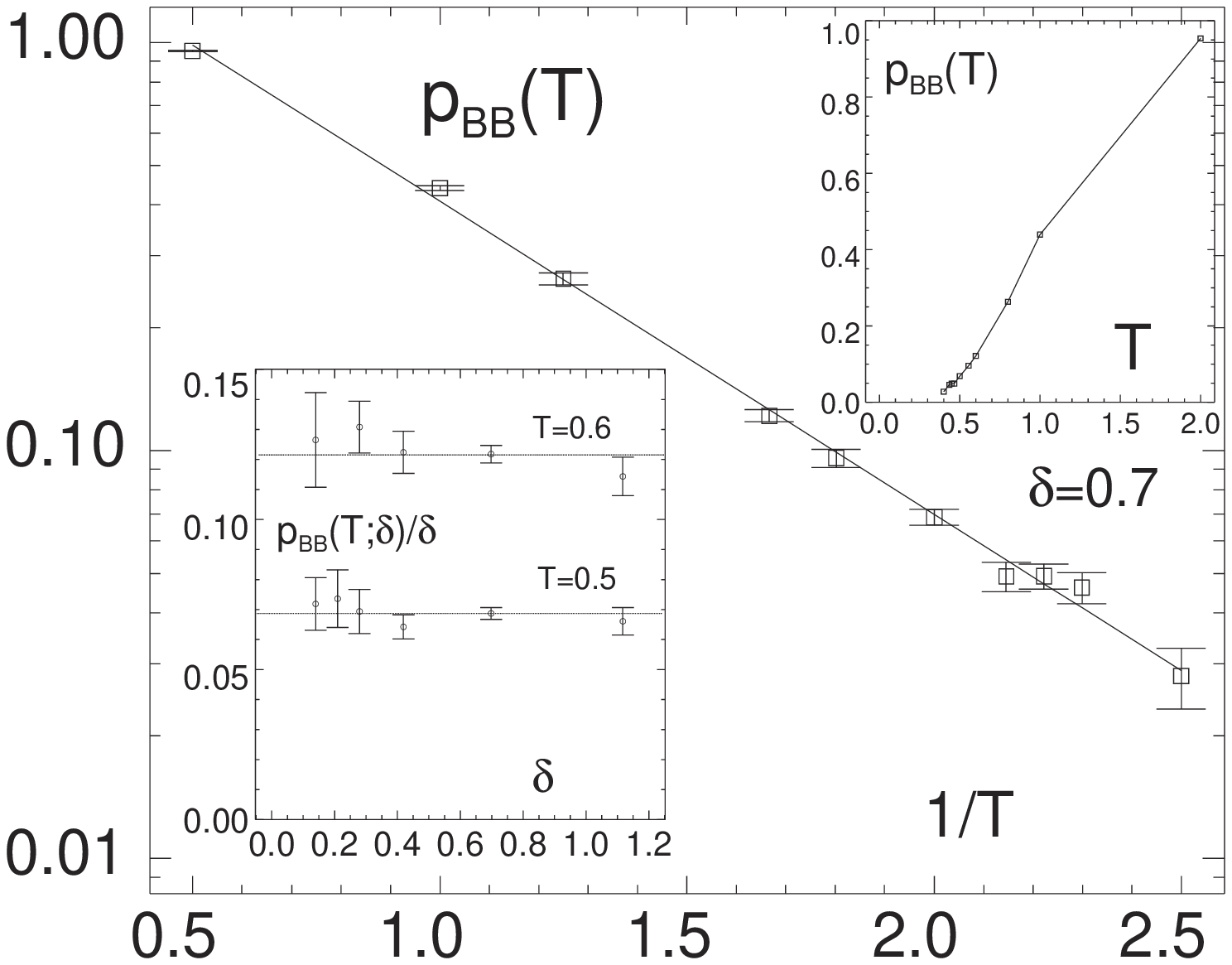}}{
Population of basin borders, $\pbb(T)$, obtained from disturbances of length
$\delta=0.7$, which corresponds to a displacement of ca.~$0.09$ per particle.
Left inset: dependence on $\delta$ of $\pbb(T)$, for $T=0.5$ and $T=0.6$.
Right inset: $\pbb(T)$ plotted linearly against $T$.
}{FIGPBB}
One obtains the behavior
\NEQU{\pbb(T;\delta)\to\const\times\pbb(T)\delta,\quad
\delta\to0}{EQUPBB3}   (the constant is set to unity for
convenience). The validity of \equ{EQUPBB3} is demonstrated in the
left inset of \figref{FIGPBB}, where $\pbb(T;\delta)/\delta$ has
been calculated as a function of $\delta$. We find that
$\pbb(T;\delta)/\delta$ is constant within statistical error below
$\delta=1.2$. As an orientation, the typical distance between
neighboring minima is larger than $2.0$, whereas intra-MB
neighbors on average are less than $1.0$ apart.

The main part of \figref{FIGPBB} shows results for $\pbb(T)$ in an Arrhenius plot, with $\delta=0.7$.
Over the whole temperature range considered, $\pbb(T)$ is Arrhenius-like.
The apparent activation energy is ca. 1.8, which is small in comparison with the typical values
observed for MB lifetimes. However, the temperature dependence becomes stronger
if we impose the constraint of a minimum distance between neighbouring minima (data not shown).
In this way, we eliminate the fast intra-MB transitions, which have small barriers.

In any event, $\pbb(T)$ features no noticeable change in behavior
when approaching and crossing $T_c$.
In a different graphical representation (see right inset) one might wrongly conclude that
$\pbb(T)$ disappears at some finite temperature.
Stated differently, the data
suggest that the increasing timescale separation upon cooling
happens rather smoothly, with no distinctly new physics emerging
near $T_c$. This is in qualitative agreement with the work of
Schr\o der et~al.~\cite{Schroder:88}, who use the incoherent
scattering functions from hopping dynamics $\xi(t)$ to deal with
the separation of intra- and interbasin dynamics. There, the
initial short-time decay of scattering functions (quantified by
the so-called non-ergodicity parameter)
is nothing else than a measure for the population of basin borders.


\section{Energy Barriers from PEL Topology.}
\label{SECBARRIERS}

\subsection{Return Probabilities and Metabasin Definition.}

\figany{\WiePosit}{\includegraphics[width=8cm]{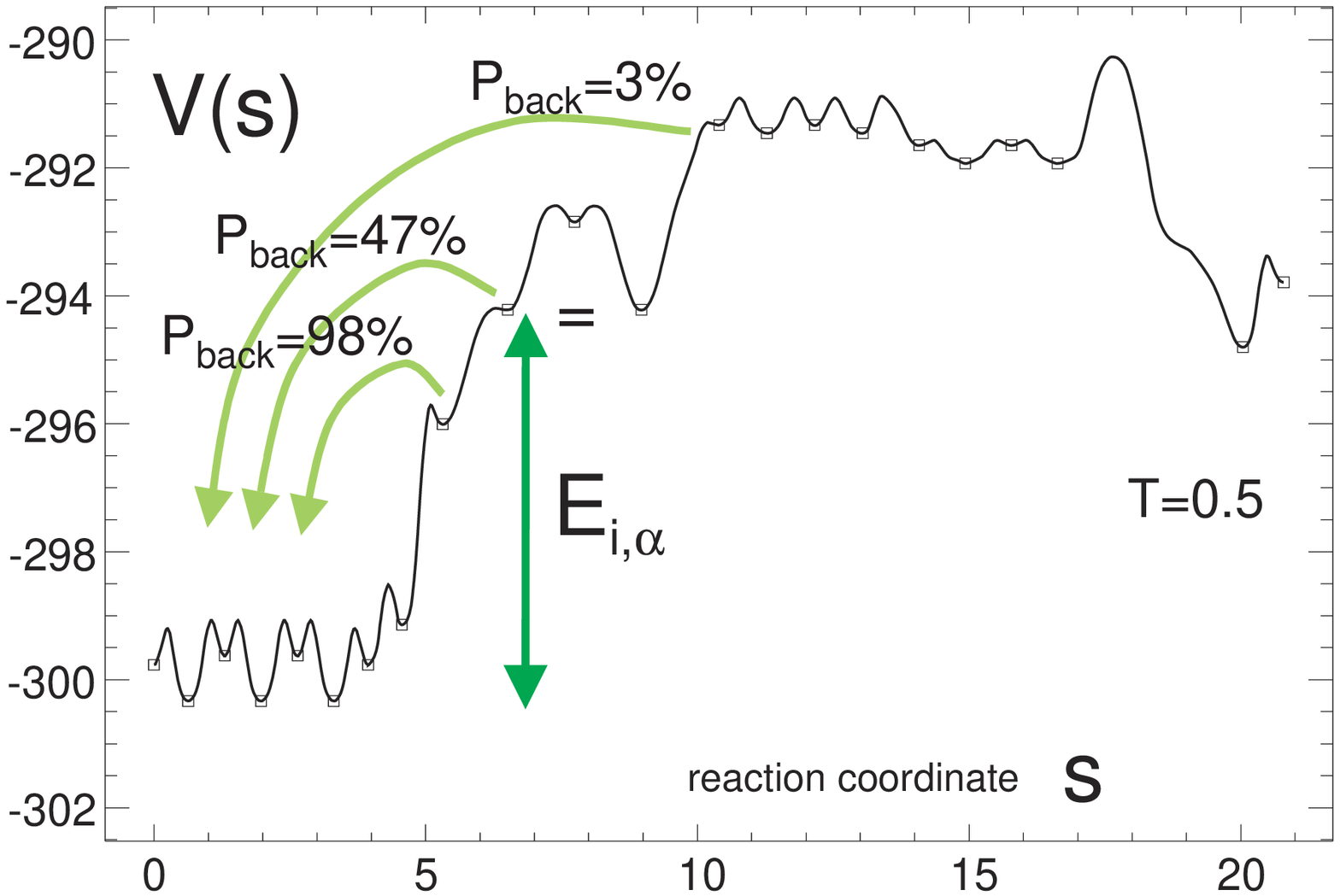}}{
Potential energy along the reaction path $\zeta(s)$, which was calculated
from the dynamics during $10^5$~MD~Steps, at the end of a typical MB of life
span $8\times10^6$~MD~Steps. The mapping of $s$ to time is non-linear.  The
small barriers for $s<5$ belong to fast intra-MB transitions.
$\pback$ denotes the probability of returning to the bottom of the MB.
As a comparison, the potential energy at that temperature ($T=0.5$)
fluctuates around $-249.3 \pm 6.1$.
}{FIGPBACK}
With the tools of interval bisection and TS search, we are now in the position to
analyze the escapes from MBs in full detail.
When a MB is left, we first resolve all minima visited during the escape.
Second, all corresponding TSs and, if desired, reaction paths are calculated.
An example is shown in \figref{FIGPBACK}.
The successive RPs were spliced together to a long, multi-minima RP $\zeta(s)$.
One might take the energy profile, $V(\zeta(s))$, depicted in the figure,
for one of the common cartoons of a PEL. However, it rests upon real data.
Berry and coworkers have produced similar charts
for the relaxation of small atomic clusters towards their global minima~\cite{Ball:379,Ball:382}.
For $s<5$ one can see the typical back-and-forth hopping
among the ground minima of the MB.
Obviously, the corresponding
barriers are not large compared to $\kB T=0.5$.
The escape starts at $s=5$.
The first minimum reached
is very unstable as expected from the small backward barrier.
Indeed, if
we repeatedly start in this minimum and perform a number of short
simulation runs   (here: $99$)  with different random numbers, the system will return to the
bottom of the MB with probability $\pback=98\%$ and leave the
range of attraction only rarely.
Thus, the escape is far from being complete at this stage.
Going to the next minimum, the return probability decreases, but does not drop
to zero.
We say that the system is free if $\pback$ is
smaller than $50\%$.
As the outcome of this investigation, we obtain
the energy barrier surmounted before the minimum with $\pback<50\%$ was reached, see below.
The exits from other long-lived MBs mostly look the same as in the
example, while the escape in one jump is not common.
In other words, MBs usually have the form of a funnel with some
ledges on the walls~\cite{Stillinger:333,Middleton:214}.
Minima with $\pback>50\%$ are said to belong to the MB.
This criterion is reminiscent of the definition of dynamic bottlenecks
introduced by Chandler and coworkers~\cite{Bolhuis:387}.

\figany{\WiePosit}{\includegraphics[width=6cm]{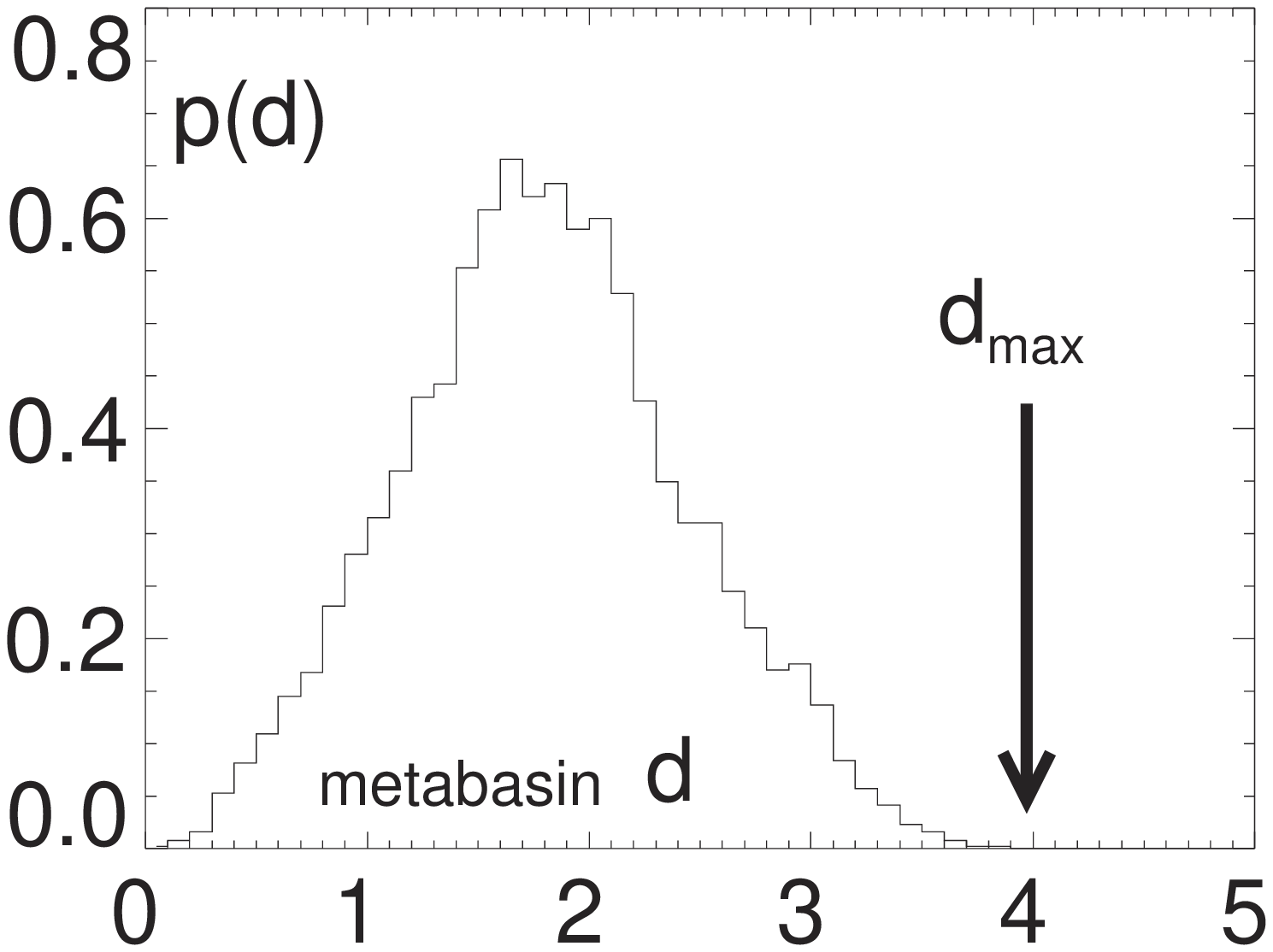}}{
Distribution of MB diameters, $d$, defined as the maximum distance
between all minima that were visited during a MB lifetime.
The delta-peak from single-minimum MBs has been omitted.
}{FIGDMAX}
An interesting property of a MB is its diameter $d$.
It is defined as the maximum distance between its minima.
For the MBs found in the simulation at $T=0.5$,
the distribution of diameters is depicted in \figref{FIGDMAX}.
The delta-peak from single-minimum MBs has been omitted.
No MB with $d>\dmax=4$ has been found.
As a consequence, if a minimum has a distance larger than $\dmax$
to some MB minimum, we can safely assume $\pback\ll 50\%$.
This criterion has already been used in section \ref{SECMTAU}.

\figany{\WiePosit}{\includegraphics[width=6cm]{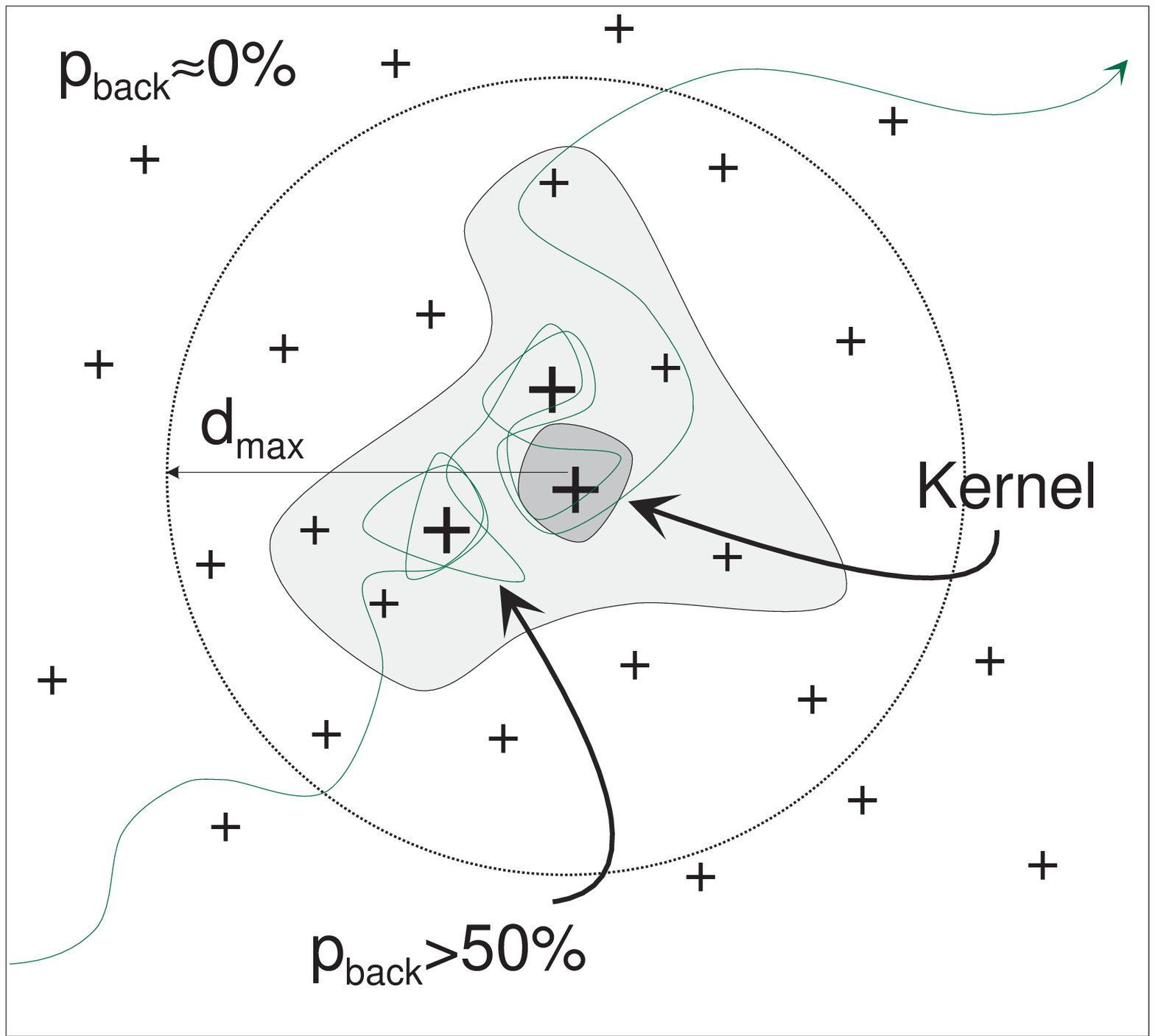}}{
Sketch of the configuration space around a MB, crosses representing minima.
Large crosses are the highly populated minima on the bottom of the MB.
The shaded area comprises minima of high return probability to the kernel minimum
($\pback>50\%$). By definition, these constitute the MB.
The bent line is the system trajectory $x(t)$ entering and finally leaving the MB.
}{FIGSCHEMAFUNNEL}
Based on these insights, we can now provide a more complete description
of MBs (\figref{FIGSCHEMAFUNNEL}).
First, the ground state of a MB has to be identified (kernel minimum),
since the definition of $\pback$ rests upon it.
At low enough temperatures, the kernel minimum will certainly be
visited during the MB lifetime, due
to the very low barriers among the minima on the bottom of the MB.
Second, for minima beyond the distance $\dmax$ from the kernel, we
set $\pback$ to zero. Third, the probability $\pback$ for
returning to the kernel before reaching a distance greater than
$\dmax$ can be assigned to every remaining minimum and -in
principle- be computed by simulation. To this end, one repeatedly
starts in the minimum and checks if a recurrence to the kernel
occurs. Fourth, the minima with $\pback>50\%$ are defined as the
MB.

Please bear in mind that $\pback$ will in general depend on temperature, since it is
defined by dynamics. Correlations among minima are expected to increase towards
lower temperatures, implying that MBs are no static concept but rather grow
with decreasing $T$. In \figref{FIGPBACK}, e.g., the minimum at $s\approx6.5$
has the 'critical' value of $\pback\approx47\%$ at $T=0.5$.
Although we do not know the details of PEL connectivity around this minimum,
the small backward barrier suggests that the minimum would exceed $\pback=50\%$
for still lower temperatures, thus joining the MB.
However, we may also conceive some situations where a critical $\pback\approx50\%$
is quite unsusceptible to temperature changes. This is the case if backward and
forward barriers are of about the same size. We will come back to that issue later.

We further note that the explicit computation of $\pback$ can be extremely expensive.
This is mainly the case when $\pback$ is small, and complete escapes
beyond $\dmax$ have to be awaited.
However, the exact value of $\pback$ is of no great interest.
In fact, it suffices to know whether $\pback<50\%$ or $\pback>50\%$.
This decision can often be reached to a high confidence with few trials.

The MB lifetime algorithm in section~\ref{SECMTAU} is based
on the detection of back-and-forth jumps between minima.
One mostly observes the  dominant minima on the bottom of the MBs,
whereas the more
elevated members are only weakly populated, see
\figref{FIGSCHEMAFUNNEL}.
If MB lifetimes are to be read from a simulation run, it suffices to notice
when the set of dominant MB minima has been left, since the visits
to the elevated minima at the end of the MB lifetime happen quite
rapidly.
Thus, the algorithm of section~\ref{SECMTAU}
reduces the MB to the most populated minima, which is sufficient for the
purpose of lifetime calculation from a given simulation run.
In contrast, for the {\it prediction} of MB relaxation behavior as pursued in this section,
the minima close to the rim of MBs are of
special interest. Their elevations from the bottom of the MB give the depth
of the MB.

\subsection{Barriers for Metabasin Relaxations.}

In the spirit of the above remarks,
we will now carry out a systematic investigation of the energy barriers overcome
when escaping MBs.
The goal to is to recover the apparent activation energies computed in section~\ref{SECMTAU}
from PEL topology.

The mean lifetime $\eww{\tau_i}$ of MB $i$ can be expressed in
terms of escape rates $\gammaia$ of different relaxation channels~$\alpha$,
\NEQU{\eww{\tau_i}^{-1}=\sum_\alpha\gammaia.}{EQUMBRATE}
In general, each $\gammaia$ reflects a multi-minima escape path
\NEQU{\xi_0\stackrel{\zeta_{01}}{\longrightarrow}\xi_1\stackrel{\zeta_{12}}{\longrightarrow}
\xi_2\ \ ...\ \ \xi_{M-1}\stackrel{\zeta_{M-1,M}}{\longrightarrow}\xi_M}{EQUJUMPSEQUENCE}
as the one shown in
\figref{FIGPBACK}.
Here, $\xi_0$ is the kernel minimum and
$\zeta_{ab}$ is the TS for $\xi_a\to\xi_b$.
Suppose that the number $M$ of jumps in the sequence \equ{EQUJUMPSEQUENCE} is large enough
to completely quit the MB's range of attraction, i.e., $\pback(M)\approx0$.
For the escape shown in \figref{FIGPBACK}, e.g., $M\ge7$ would be fine.

We further take for granted that the rates for single barrier crossings
follow quantitatively -via transition state theory- from
the height of barriers, $E_{ab}=V(\zeta_{ab})-V(\xi_a)$
(the energy difference between the minimum $\xi_a$ and the TS between $a$ and $b$).
Hence, rates $g_{ab}$ for single transitions $\xi_a\to\xi_b$ are characterized by
\NEQU{g_{ab}\propto e^{-\beta E_{ab}}.}{EQURATE}
A justification for this assumption, even for temperatures above $T_c$,
will be given in section~\ref{SECACT}.

Generally, the probability of upward jumps is small at low $T$.
Hence, climbing out of a MB in a back-and-forth fashion (e.g., $\xi_a=\xi_{a+2}$ and
$\xi_{a+1}=\xi_{a+3}$)
is not probable.
(This is reminiscent of the fact that the activated crossing of single potential barriers
happens on a short time scale, i.e. in a rather straight way.)
In contrast, {\it excursions} from the main path may happen.
As shown in \figref{FIGPBACK}, the minimum at $s=6.5$ is revisited at $s=9$ after
taking a look at another minimum ($s\approx8$).
The latter does not appear again later on.
Clearly, running into such 'dead ends' should not contribute to the escape rate via
the successful main path. We therefore eliminate such excursions
from the sequence of minima, \equ{EQUJUMPSEQUENCE}.
From these remarks we take the liberty of assuming that no minimum appears more than
once along the escape path,
\NEQU{\xi_a\ne\xi_b,\quad a\ne b.}{EQUDIFFMINS}

We are now interested in the contribution of the path \equ{EQUJUMPSEQUENCE} to the total
escape rate~\equ{EQUMBRATE}. Particularly, we have to consider the question
of how many single transitions are relevant for the escape process.
The probability to jump from minimum $\xi_a$ to $\xi_{a+1}$ is $g_{a,a+1}/g_a$,
where $g_a$ denotes the inverse lifetime of minimum $\xi_a$.
The rate of escape via a longer pathway now is given by the rate of the first jump
times the probability that the minima $\xi_a$ ($a=1,...M$) are visited in correct order thereafter,
\NEQU{\gammaia=g_{01}\frac{g_{12}}{g_{1}}\frac{g_{23}}{g_{2}}...
\frac{g_{M-1,M}}{g_{M-1}}.}{EQUGAMMAESCAPE}
With the help of \equ{EQURATE} one calculates
\NEQU{-\frac{\d}{\d\beta}\ln\gammaia
=E_{01}+\sum_{a=1}^{M-1}\pret(a)(E_{a,a+1}-E_{a,a-1}),
}{EQUBARRIER1}
where $\pret(a)=g_{a,a-1}/g_a$ is the probability to jump back to minimum~$a-1$ from minimum~$a$.
In the derivation of \equ{EQUBARRIER1}, we have neglected a term proportional
to $E_{a,a+1}$ minus the average barrier when jumping from~$a$ to a neighbouring
minimum other than~$a-1$.
This term strictly vanishes when performing the final summation in \equ{EQUMBRATE}.
Moreover, we made use of \equ{EQUDIFFMINS}.

One possibility for calculating activation energies from \equ{EQUBARRIER1}
would be to consider the complete paths \equ{EQUJUMPSEQUENCE}, where $\pback(M)\approx0$,
and determine all terms in the sum of \equ{EQUBARRIER1}.
However, an accurate computation of
all the desired $\pret(a)$'s would even be more costly than the determination of
the point where $\pback$ changes from above to below $50\%$.
We therefore use the following approximation of \equ{EQUBARRIER1}, which is in
conformance with our previous definition of MBs:
Let $m(T)$ be the first minimum along the path \equ{EQUJUMPSEQUENCE}, where $\pback<50\%$.
Then, for all $a<m(T)$, we set $\pret(a)$ to unity, while for $a\ge m(T)$
(i.e. outside the MB), we let $\pret(a)=0$.
Thus,
\NEQU{\SPLIT{
-\frac{\d}{\d\beta}\ln\gammaia
\approx \Eia&\equiv E_{01}+\sum_{a=1}^{m-1}(E_{a,a+1}-E_{a,a-1})\\
&=\eps_{m-1}-\eps_0+E_{m,m-1},
}}{EQUBARRIER}
where $m=m(T)$.
In this way, the terms $a<m(T)$ in \equ{EQUBARRIER1} are given higher weights,
whereas those of $a\ge m(T)$ are neglected.
We will dwell on the quality of this approximation later on.

Note that, due to the temperature dependence of $\pback$, energy barriers $\Eia$
generally increase upon cooling:
At high temperatures, in contrast, correlations among minima are small, such that
MBs (even the low-lying) consist of only one minimum.
This effect is included in \equ{EQUBARRIER} by the temperature dependence of $m(T)$.

\subsection{Single Metabasins.}
We now relate the lifetimes of single, selected MBs (cf.
section~\ref{SECMTAUFUNNEL}) to PEL barriers. By repeated starts
from these MBs, the local PEL topography is sampled thoroughly,
yielding sets of typical escape pathways. Whenever a MB is left,
we locate the transitions by interval bisection and obtain the
corresponding TSs with the help of the ridge method. Then,
$\pback$ is calculated for the minima visited, until for the first
time, $\pback<50\%$. Finally, the barrier $\Eik$ is computed
according to \equ{EQUBARRIER}, where $\alpha(k)$ denotes the
escape path chosen at the $k$th escape.
\figany{\WiePosit}{\includegraphics[width=8cm]{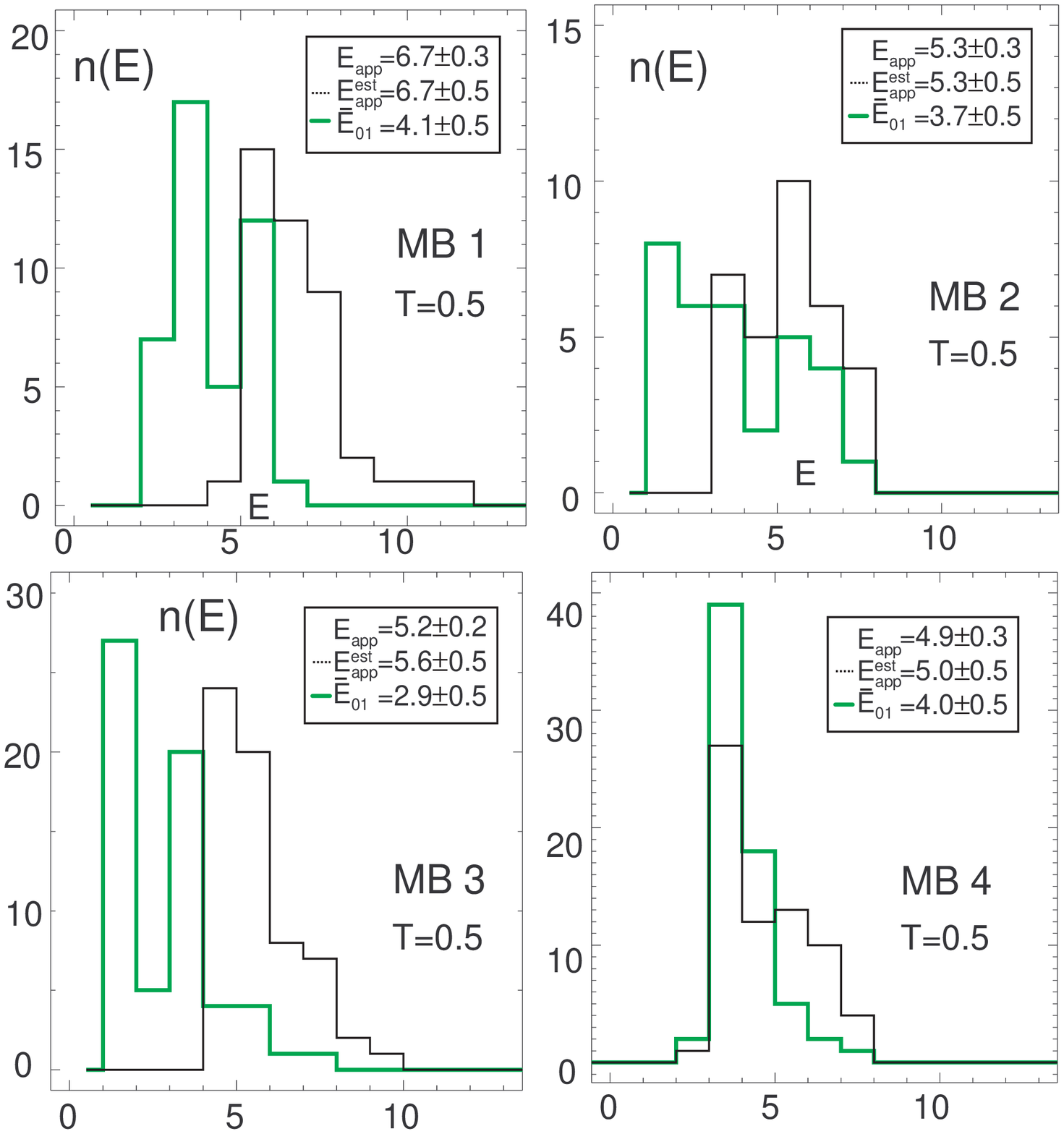}}{
Bold curves: Histograms of barriers $\Eik$ overcome when escaping single
MBs ($i=1,2,3,4$ at $T=0.5$).
Light curves: Respective histograms of barriers $E_{01}$ from first jumps.
Apparent activation energies $\Eapp(i)$, mean barriers $\Eappest(i)$, and
mean barriers from first jumps $\Efst$ are given in the figure.
}{FIGPOFEFUNNEL}
The histograms of barriers are shown in \figref{FIGPOFEFUNNEL},
for the four MBs of \figref{FIGTAUFUNNEL}, at $T=0.5=1.1T_c$.
Due to the slow dynamics at this temperature, the computation of $\pback$
was rather expensive.
Nevertheless, the statistics should be sufficient for
a reasonable estimate of the apparent activation energy.
To this end, we express $\Eapp(i)$ of MB $i$ in terms of the contributions $\Eia$,
\NEQU{\frac{\d}{\d\beta}\ln\eww{\tau_i}\approx\eww{\tau_i}\sum_\alpha\Eia\gammaia
=\sum_\alpha\pia\Eia\equiv\Eappest(i),}{EQUEAPPI}
where \equtwo{EQUMBRATE}{EQUBARRIER} were used.
Thus, the barriers $\Eia$ are weighted by the probabilities
$\pia=\gammaia/\sum_\alpha\gammaia$ that the escape happens
via pathway~$\alpha$.
Note that the $\Eik$ correspond to the pathways that were
{\it chosen} by the system, i.e. they are already weighted correctly
by $\pik$, compare \equ{EQUEAPPI}.
Therefore, $\Eappest(i)$ is just the average of the $\Eik$.
The values of $\Eapp(i)$ and $\Eappest(i)$, given in \figref{FIGPOFEFUNNEL},
are in good agreement.
Also shown in \figref{FIGPOFEFUNNEL} is the distribution of first barriers, $E_{01}$,
belonging to the step $\xi_0\to\xi_1$.
Evidently, the neglect of the multi-minima nature of escapes leads to a considerable
underestimation of apparent activation energies.

\figany{\WiePosit}{\includegraphics[width=7cm]{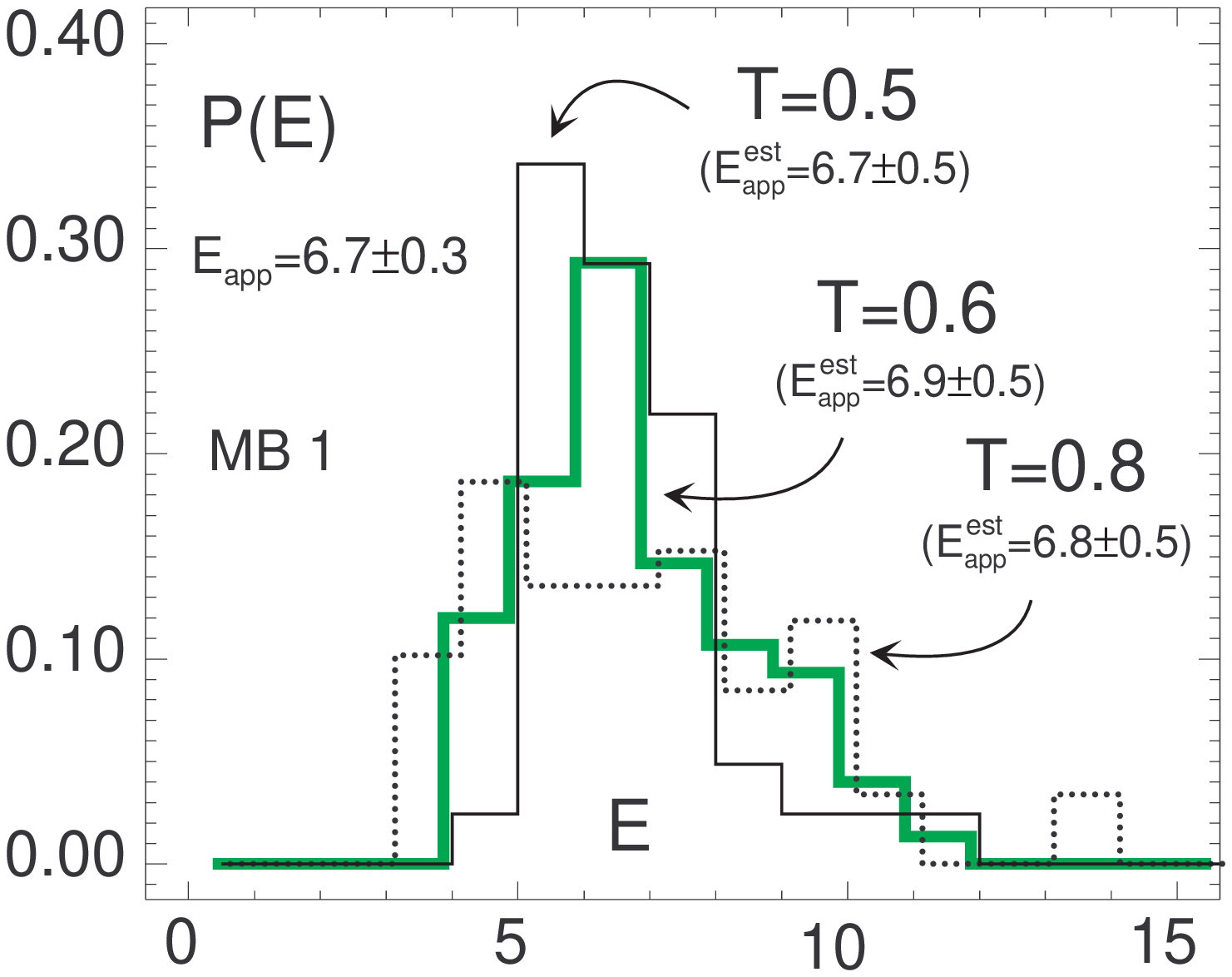}}{
Normalized histograms of barriers $E_{1,k}$ overcome when escaping MB 1, for
$T=0.5,0.6,$ and $0.8$. The number of contributing barriers are 42, 72, and 59,
respectively.
Estimated apparent activation energies, $\Eappest(i;T)$, are given in the figure.
}{FIGACT20}
We now continue the discussion of the temperature dependence of barriers $\Eia(T)$.
At the example of MB~1 from \figref{FIGPOFEFUNNEL}, we have
carried out the above program for two other temperatures,
$T=0.6$ and $0.8$. The obtained distributions of barriers, $P(\Eia)$,
are shown in \figref{FIGACT20}.
We find that the estimates for the apparent activation energy
($\Eappest(1)=6.9\pm0.5$, $T=0.6$, and $\Eappest=6.8\pm0.5$, $T=0.8$) remain
in good agreement with $\Eapp(1)=6.7\pm0.3$ from section~\ref{SECMTAU}.
The distributions of barriers, however, grow narrower with decreasing temperature.
Single, high barriers, contributing to the right wing of the distribution,
become inaccessible at low $T$, i.e., the relative weights $\pia$ of the corresponding
escapes become small. This suppression of high barriers at low $T$ is a trivial effect.

More interesting is the vanishing of small barriers upon cooling, i.e.,
of the barriers $E<5$ in the figure.
Naively, one would expect these to dominate the escape rate at low $T$.
However, due to the stronger backward correlations (increased $\pback$),
jumps over these barriers eventually do not suffice anymore to escape.
As described above, the respective escape paths, $\xi_0\rar...\rar\xi_{m(T)}$, grow longer,
and the barriers change to a different, mostly larger value.

\subsection{Average over Metabasins.}
\figany{\WiePosit}{\includegraphics[width=7cm]{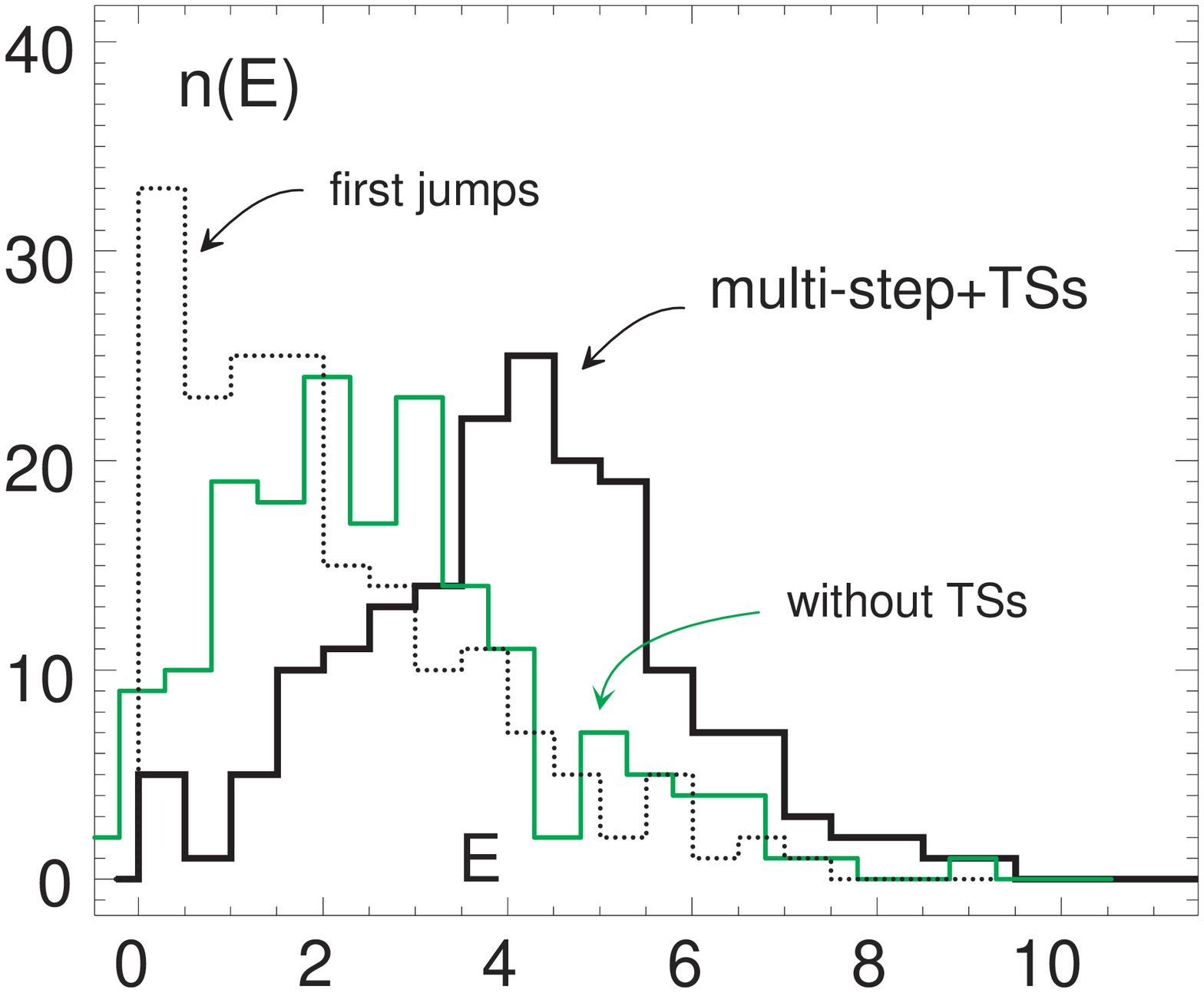}}{
Histogram of barriers from a regular MD run at $T=0.5$ (bold).
Neglecting the contributions of the last transition state ($\eps_m-\epsmb$),
we find smaller barriers (light line).
The barriers $E_{01}$ from only the first jumps are given as the dotted line.
}{FIGPOFE}
During our analysis of the escape times in section~\ref{SECMTAU}
the apparent activation energies $\Eapp(\epsmb)$ emerged as
useful quantities. Although the above results
already indicate that barrier hopping is the relevant motional
mechanism, a clear-cut verification requires the comparison with the
average barrier the system has to cross when leaving a MB with
energy $\epsmb$.

For this purpose we now carry out a similar program as before, with
many MBs visited during an ordinary MD run. We concentrate on MBs
with lifetimes of more than $10^5$~MD~steps (179~MBs) at $T=0.5$.
When such a MB is left, we locate the transitions by interval
bisection and obtain the corresponding TSs by the ridge method.
Then, we calculate $\pback$ and identify the barrier
$E_k\equiv\Eikk$ according to \equ{EQUBARRIER}.
The histogram of barriers is shown as the bold line in \figref{FIGPOFE}.
For comparison, we also show the barriers minus the contribution of
the TSs $E_{(m-1)m}$. Ignoring multi-minima correlations, we further
show the histogram of first barriers $E_{01}$ of escapes.
Evidently, the neglect of TSs or of backward correlations leads to much smaller
barriers.

From the above barriers,
we will now calculate estimates of apparent activation energies.
When the average over lifetimes of different MBs is considered,
each MB $i$ acquires a weight $\phi_i$ corresponding to its
probability of occurrence,
\EQU{\tauphi=\sum_i\phi_i\eww{\tau_i}.}{}
At fixed $\epsmb$, the analog to
\equ{EQUEAPPI} can then be derived
\NEQU{\frac{\d}{\d\beta}\ln\tauepsmbT
\approx\sum_i\frac{\eww{\tau_i}\phi_i}{\tauepsmbT}\sum_\alpha\pia\Eia,
}{EQUEAPPXX}
where summation goes over MBs of energy $\epsmb$.
As in \equ{EQUEAPPI}, the barriers in \equ{EQUEAPPXX}
are weighted according to their
probability of occurrence, but, additionally,
with the respective MB lifetimes.

In \equ{EQUEAPPXX}, we have neglected terms stemming
from the variation of $\phi_i$'s with temperature.
This is justified, since the $\phi_i$'s belong to the same $\epsmb$.
Their relative weights will
only vary if these MBs differ considerably in barrier heights.
As already stated above, however, MBs of the same energy seem to be fairly
uniform regarding this property.
For the finite sample of MBs visited during an MD run,
\equ{EQUEAPPXX} then takes the form
\NEQU{\Eappest(\epsmb)=\frac{\sum\tau_k E_k}{\sum\tau_k},}{EQUEAPPSIM}
where summation goes over MBs of energy $\epsmb$.
Again, the correct weighting is implicit here. This expression can
be shown to converge to the right-hand side of \equ{EQUEAPPXX}
in the limit of infinitely long sampling.
\figany{\WiePosit}{\includegraphics[width=7cm]{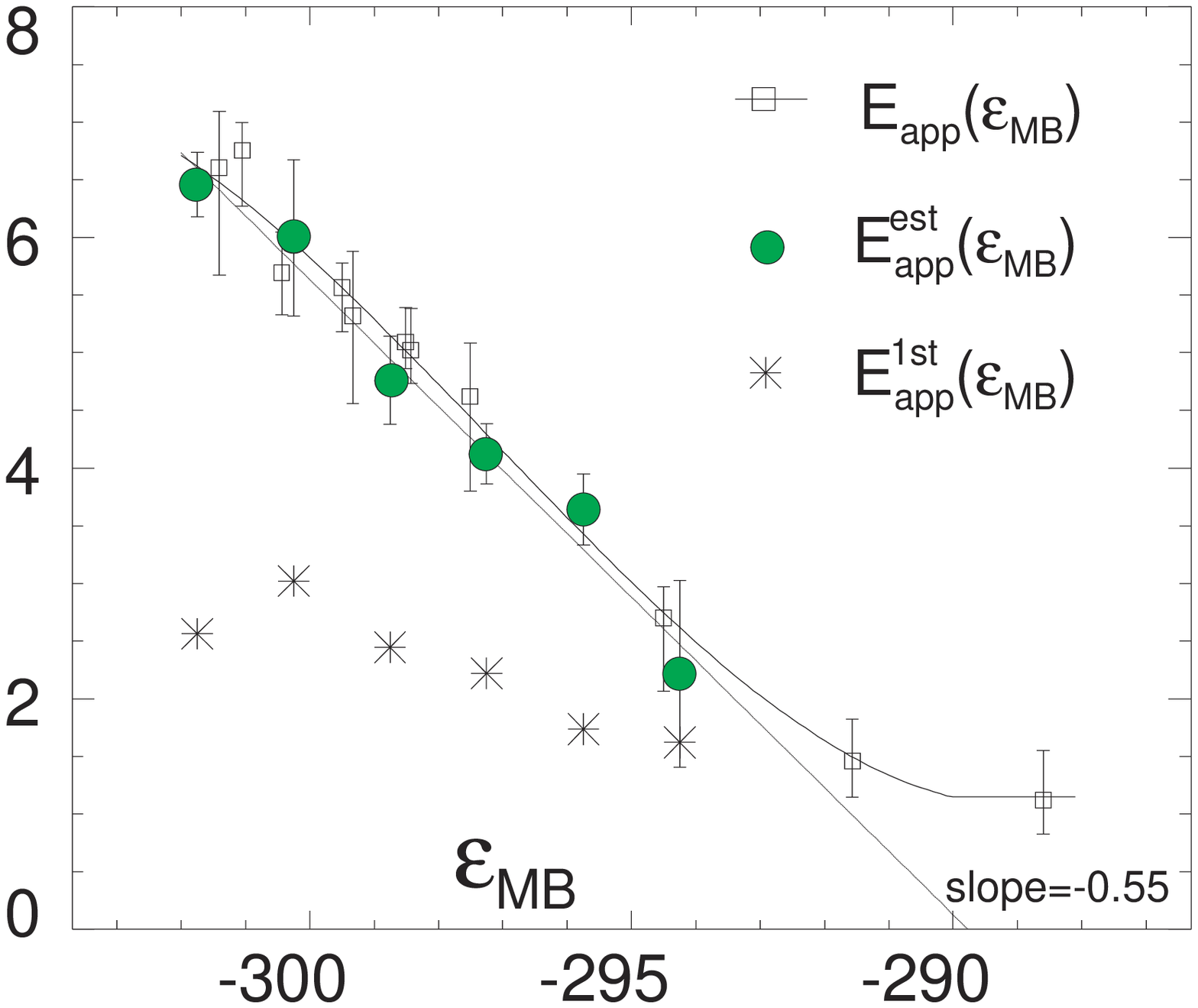}}{
$\Eapp(\epsmb)$ (\figref{FIGTAU}) vs. estimated $\Eappest(\epsmb)$
from PEL barriers.
Considering only the first jumps of escapes, we find a much smaller estimate
($E_\ind{app}^\ind{1st}(\epsmb)$).
Data stem from a regular MD run at $T=0.5$, where MBs of lifetime greater
than $10^5$~MD~steps were analyzed (179~MBs, see \figref{FIGPOFE}).
}{FIGEAPP}
In \figref{FIGEAPP} we show the values of $\Eappest(\epsmb)$, determined in
this way. They perfectly agree with the apparent activation
energies, derived from the analysis of relaxation times at
different temperatures. Thus we have a clear-cut proof that the
apparent activation energies $\Eapp(\epsmb)$ are indeed related
to barriers on the PEL and thus reflect activated behavior
significantly above $T_c$. This again demonstrates that we not only
deal with the right order of barrier sizes, but we also {\it
quantitatively} link PEL topography to dynamics.

For comparison, we included the apparent activation energy which
results, if only the first transitions of escapes, $\xi_0\to\xi_1$,
are considered ($E_{01}=V(\zeta_{01})-\eps_0$).  One ends up with
much too small apparent activation energies.  Again, multi-minima
correlations turn out to be crucial for the characterization of
MB depths.

In principle, the results of \figref{FIGEAPP} may slightly change if all MBs
rather than those with lifetimes larger $10^5$~MD~steps were
considered. However, our analysis has clearly revealed (see, e.g.,
\figref{FIGTAUFUNNEL}) that the depth of the trap only mildly varies when
comparing MBs with similar $\epsmb$. Thus inclusion of MBs with
smaller values of $\tau$ would not significantly change the values
of the apparent activation energies $\Eappest(\epsmb)$.

\figany{\WiePosit}{\includegraphics[width=8cm]{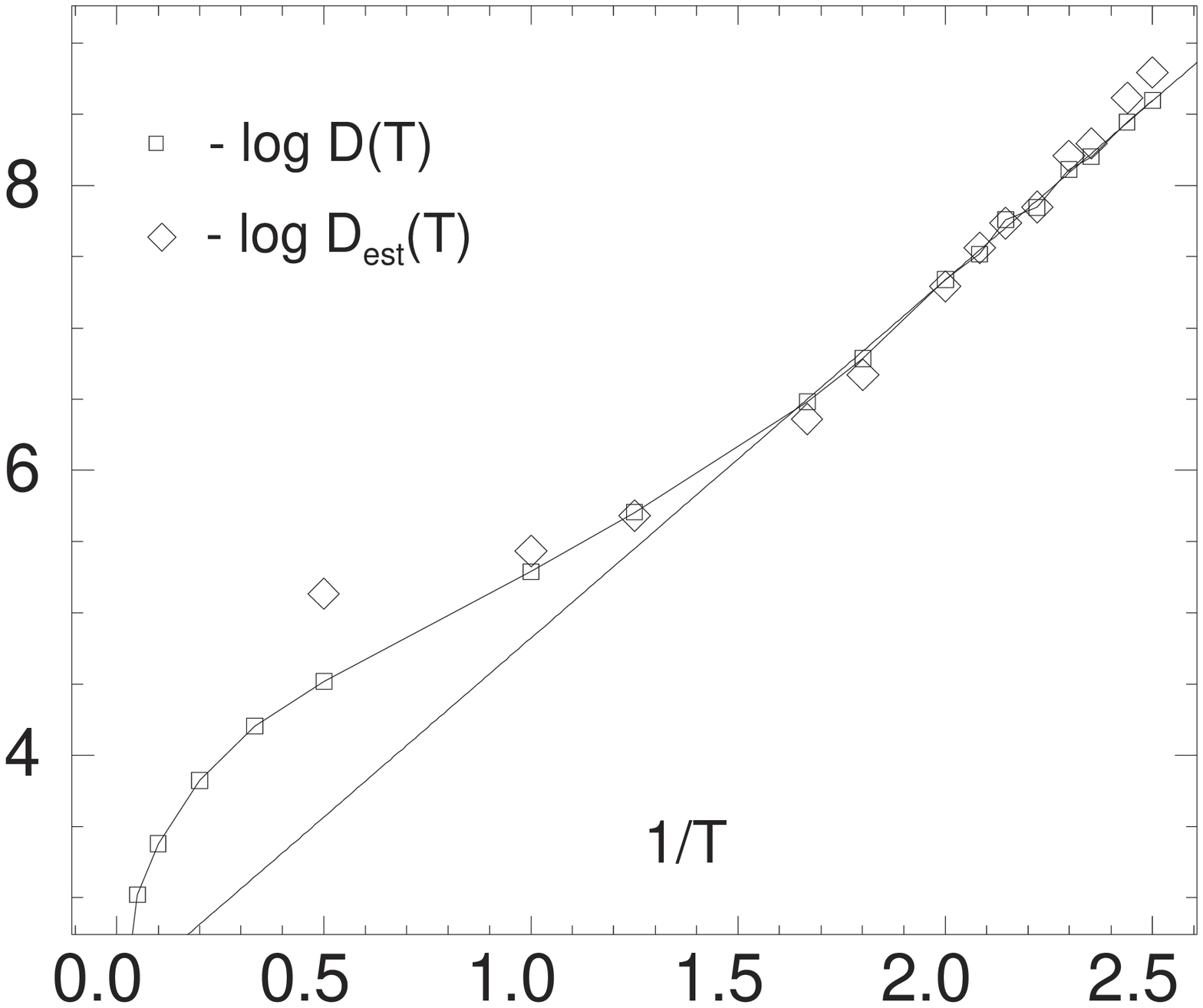}}{
Comparison of the inverse diffusion constant, $1/D(T)$, with
the prediction $1/D_\ind{est}(T)$ from \equ{EQUDIFFTDPEL2},
$\tau_0=200$.
}{FIGDIFFTAU}
Finally, we show that these results, in conjunction with $p(\epsmb;T)$,
largely explain the behavior of the diffusion constant $D(T)$.
This is a conceptually important step, since we link $D(T)$
to purely structural and thermodynamical quantitites, see \equ{EQUCONCEPT}.
The key is the mean lifetime $\tauepsmbT$ of MBs at energy $\epsmb$,
which is parametrized by $\tau_0(\epsmb)$ and $\Eapp(\epsmb)$ (\equ{EQUTAUEPSMB}).
The former, $\tau_0(\epsmb)$, however, has not been deduced from PEL properties.
Its variation with MB energy is not strong (\figref{FIGTAU}(c)),
so we can hope that setting it to a constant will be a good approximation.
Thus, \equ{EQUDIFFTDPEL} becomes
\NEQU{D(T)\approx
\frac{a^2}{6N\tau_0}\int\d\epsmb p(\epsmb;T)e^{-\beta\Eappest(\epsmb)}
\equiv D_\ind{est}(T).
}{EQUDIFFTDPEL2}
The estimated diffusion constant derived from this expression is shown in \figref{FIGDIFFTAU}.
The agreement of $D(T)$ with our estimate is satisfactory below $T=1$, albeit we find
a slightly too strong temperature dependence for the lowest $T$.
The deviation at $T=2$ is due to the depart of $\tauepsmbT$ from Arrhenius behavior, see \figref{FIGTAU}(a).


\section{Barrier Crossing.}
\label{SECACT}

When making use of \equ{EQURATE}, we presumed that the barriers
$V(\zeta_{ab})-V(\xi_a)$ in fact are the determinants of the
temperature dependence of rates.
The excellent agreement between
$\Eapp(\epsmb)$, determined from dynamics, and the
$\Eappest(\epsmb)$, from the analysis of the PEL,
strongly indicates that this
presumption is indeed true. We will show here in a very detailed way
that at $T=0.5=1.1T_c$, escapes from stable MBs are perfectly
activated. More precisely, two conditions are fulfilled, (i) the
potential barriers are much larger than $\kB T$, (ii) rates follow
from the 1D energy profile of the RP plus corrections from
perpendicular curvatures.

We will check these conditions explicitly here by an analysis of
escape dynamics out of MBs. We made the observation that during every escape
from a stable MB, at least one single barrier larger than $6\kB T$
must be surmounted. Moreover, this larger jump is mostly undertaken
from one of the lowest minima of the MB, compare \figref{FIGPBACK}.
From the repeated escape runs of section~\ref{SECMTAUFUNNEL}, we
selected the most frequent ten transitions of that kind. From the
respective TSs, $\zeta_{l}$, we computed the RPs, denoted
$\zeta_l(s)$, $l=1...10$.
We then investigated the motion within the MBs over a long period of the simulation
where no escape had happened ($10^7$~MD~steps each MB).
The goal was to observe how the system tries to climb the different RPs.
To this end, we projected the instantaneous configuration $x(t)$
onto each of the RPs, according to
\EQU{s_l(t)\equiv\left\{ s': ||x(t)-\zeta_l(s')||=\min_s||x(t)-\zeta_l(s)|| \right\},}{}
which means the point on the RP next to $x(t)$.
Due to the long residences in the MBs, motion therein is largely equilibrated.
Hence, if the potential energy profiles $V\left(\zeta_l(s_l)\right)$ along
the reaction paths are of importance for the transition rates, we
expect that the populations $p_l(s_l)$ of the RPs follow from
Boltzmann's law
\EQU{p_l(s_l)\propto\exp\left\{-\beta V_l(s_l)\right\}Y_l^\perp(s_l)
\equiv\exp\left\{-\beta F_l(s_l)\right \}.}{}
The vibrations perpendicular to path $l$ are
accounted for by the harmonic partition function
\EQU{Y_l^\perp(s_l)=\int\d x \exp\brac{-\frac{\beta}{2}\sum\lambda_\nu x_\nu^2}\delta(x\cdot\hat t(s_l)),
}{}
where $\lambda_\nu$ are the eigenvalues of the Hessian matrix, $H(s_l)$,
$x_\nu$ the components of $x$ along the eigenvectors,
and $\hat t(s_l)$ is the tangent to the reaction path.

\figany{\WiePosit}{\includegraphics[width=6cm]{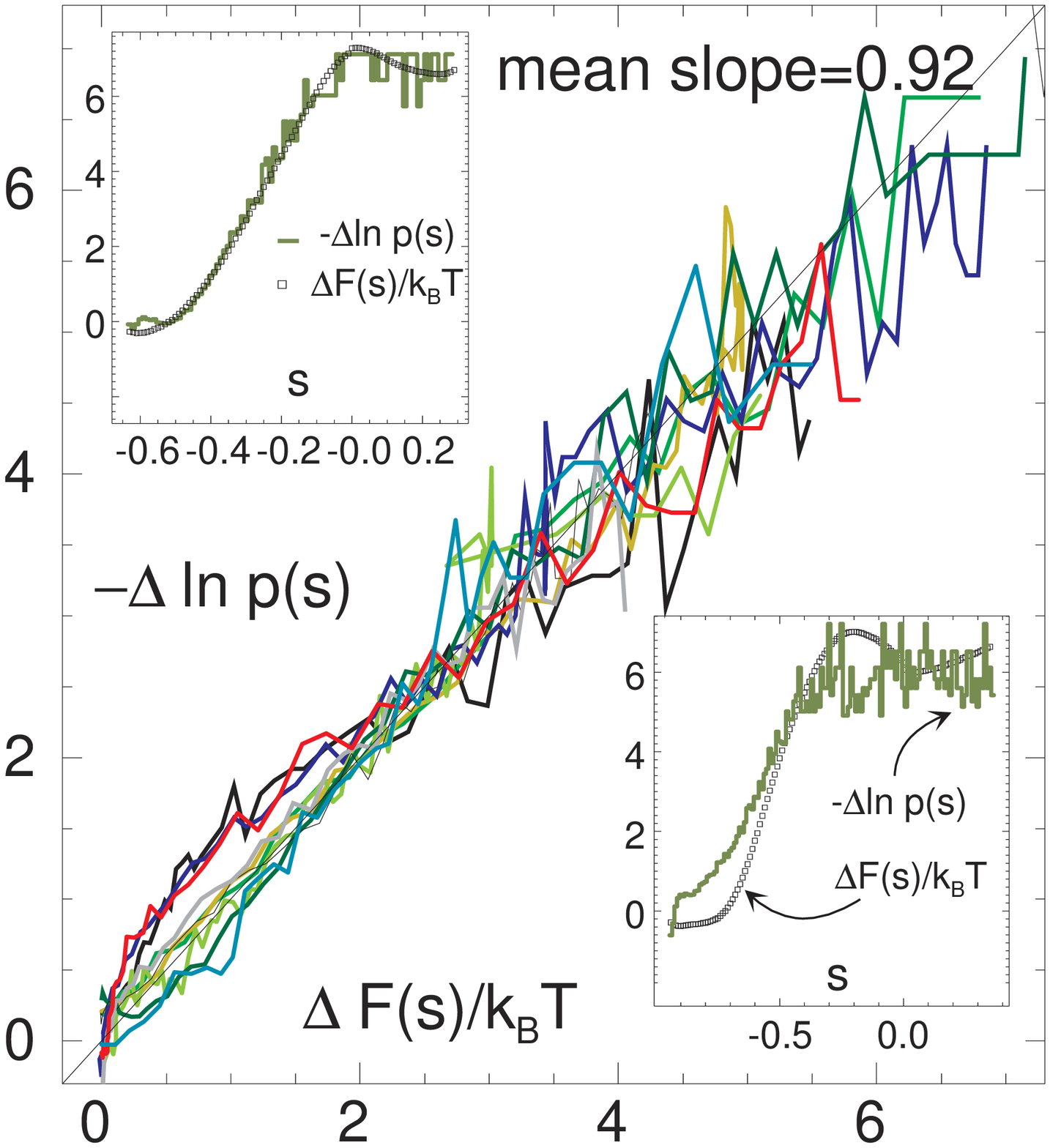}}{
Parametric plot showing the correspondence of $-\ln
p^l(s_l)+\rm{\const.}$ to the free energy profile $F_l(s_l)/\kB T+\rm{const.}$,
$l=1...10$, $T=0.5$.
All curves were shifted to start in the origin.
Insets: comparison of the free energy profiles of
two reaction paths with the population along the
path.
}{FIGACT}
The upper inset of \figref{FIGACT} shows an example of
$p_l(s_l)$ vs. $F_l(s_l)/\kB T$.
The population of the reaction path
follows nicely the prediction from its energy profile.
For RPs with complicated shapes, this correspondence
can be disturbed.
The worst agreement of the considered RPs is shown in the second inset.
Still, a clear correlation of RP population with energy is present.
We compiled the results for all 10 RPs in \figref{FIGACT}
as a parametric plot of $-\ln p_l$ \vs $F_l/\kB T$.
Curves of slope one result from a perfect equivalence of $p_l$ to $F_l/\kB T$.
Here, we find an average slope of $0.92$.
Since transition rates are proportional to the population of TSs,
the implications of these results are obvious:
MB jump rates follow from energy barriers.
We finally note that the vibrational terms $\ln Y^\perp_l(s_l)$ are minor as compared to
$\beta V(s_l)$.

In view of these results, it is a little surprising that the TS location with the help of the
auxiliary potential $\tilde V$ was that unsuccessful (cf. section~\ref{SECNLRM}).
Since the RP population suits well the harmonic description of the RP, one expects
that motion near the TS is quite harmonic, too.
Minimizing $\tilde V$ in a harmonic potential directly yields the stationary state.
Consequently, one should easily find the TS when starting from a configuration at $s\approx0$.
After section~\ref{SECNLRM}, this is not the case, so at least minor anharmonicities
must be present.


\section{Discussion.}
\label{SECDISCUSSION}

The metabasin concept is at the heart of the present study.
The important insight is that, upon cooling,
not only the time scale of inter-{\it basin} transitions becomes well separated
from intra-{\it basin} vibrations,
but also that a similar separation occurs between
{\it MB} hopping and intra-{\it MB} transitions.
Recently, Biroli and Kurchan have analysed the general problem of
defining metastable states in glassy systems~\cite{Biroli:390}. They conclude that one has
no absolute notion of a state without making reference to a time scale
and hence to dynamics. Also the present definition of MBs relies on the
dynamics of the system. It is, however, independent of time scale and
exclusively depends on the $\pback$ values which directly reflect the
topological properties of the PEL.

Our MB definition (see \figref{FIGSCHEMAFUNNEL}) is devised
to eliminate the information on trivial back-and-forth jumps within MBs.
This strongly correlated type of motion is reminiscent of the particles' rattling
in the cages formed by their neighbors.
Similarly, escaping from MBs seems to be equivalent to the breaking of cages and thus
to structural relaxation.
Guided by this idea,
we have examined MB relaxation in great detail:

First, for repeated relaxation from the same MB, we calculated the
mean relaxation time $\eww{\tau_i}$ and found Arrhenius behavior
in all cases.
The simplest view is that the apparent activation energies
$\Eapp(i)$ from the Arrhenius-like $\eww{\tau_i}$ (\figref{FIGTAUFUNNEL})
should correspond to the depths of these MBs, i.e. to the typical heights of barriers
that surround the MBs.
Indeed, this has been quantitatively confirmed for the four
randomly selected, low-lying MBs (see \figref{FIGPOFEFUNNEL}).
A direct conclusion from the constancy of $\Eapp(i)$ is that the system does
not find smaller and ever smaller barriers upon decreasing $T$.

Although not of statistical relevance for the whole PEL,
the results for the four single MBs
give us a detailed picture of the local PEL topography.
An important outcome is the variation of barrier heights with temperature,
see \figref{FIGACT20}.
We have already discussed that low barriers
increase upon cooling, due to enhanced multi-minima correlations (growing MBs),
while unnecessarily high barriers are suppressed.
Both effects seem to cancel, so that the mean barrier, $\Eappest$,
remains constant, leading
to Arrhenius behavior below $T=1$.
This cancellation may be fortunate, at least we can offer no explanation for it, here.
As depicted in \figref{FIGACT20}, the distribution of barriers becomes more and more
narrow when going from $T=0.8$ to $T=0.5$, but the mean value, i.e. $\Eappest(1)$,
remains constant.
The constant apparent activation energy of MB~1 down to $T=0.45$ implies that
the mean value of the distribution of barriers has not increased.
We thus speculate that the growth of barriers due to increasing
multi-minima correlations has essentially come to an end at $T\le0.5$.
Although the temperature dependence of the barrier distribution
has only been analyzed for a single MB, the constancy of apparent activation energies
of the other three MBs and the temperature independence of $\Eapp(\epsmb)$ support this
idea.
Stated differently, the development of superstructures of minima seems to cease
at some temperature above $T_g$.
Expressed by $\pback$, this means that no minimum with $\pback<50\%$ will surpass
$\pback=50\%$ upon further cooling, thus being unable to join the MB in question.
Hence, an escape sequence found at one temperature $T\le0.5$ has the same length at another one,
i.e., from some temperature on, the minimum $\xi_{m(T)}$
remains at $\pback<50\%$ for $T\to0$;
we then say it terminates the sequence.
It is an interesting question under what circumstances
such termination happens.
A trivial example would be a 'transit' minimum with one backward
and one forward exit, where taking the forward leads to a minimum with $\pback\approx0$.
If the backward barrier was higher than the forward one, $\pback$ would go to zero
for $T\to 0$.
On the other hand,
the minima inside MBs generally feature growing $\pback$'s upon cooling, because
the energetic gain of returning becomes more and more attractive.
Ideally, thus, for $T\to0$, we would have $\pback\to1$ within MBs, and $\pback\to 0$
outside.
This provides a plausible, physical basis for computing barrier heights according
to \equ{EQUBARRIER}, at least in the limit $T\to0$.
Clearly, a more detailed investigation of the temperature dependence of $\pback$
is necessary to back these conclusions.

Second, we analyzed the average relaxation times $\tauepsmbT$ from MBs at fixed energy $\epsmb$.
Again, they displayed Arrhenius behavior, with apparent activation energy $\Eapp(\epsmb)$
(see \figref{FIGTAU}),
which compared well with the prediction from PEL barriers (\figref{FIGEAPP}).
In this connection, a recent paper by Grigera~et~al.~\cite{Grigera:264}
is of interest.
The authors use the $\tilde V$-potential to compute
saddles in a binary soft-sphere mixture ($N=70$).
From the TSs among these saddles (index one, no shoulder), they perform
steepest descents to obtain the connected minima.
They define barriers as the energy difference $\Delta U$ from the TSs to the lower
one of the connected minima $\eps=\min(\eps_0,\eps_1)$.
Plotting the average $\overline{\Delta U}(\eps)$, they find a similar curve to our
$\Eapp(\epsmb)$, \figref{FIGEAPP}, i.e., a strong increase of barriers towards lower
energies.
In contrast, when carrying out the same analysis for
our BMLJ65,
we found a nearly constant $\overline{\Delta U}(\eps)$,
a curve close to the first barriers of escapes $E_{01}$
shown in \figref{FIGEAPP}.
We would have expected this result,
since the multi-step nature of escapes
in the BMLJ65 has clearly been demonstrated.
On the other hand, the contrasting result of Grigera~et~al.
indicates that
the soft-sphere PEL is not organized in multi-minima superstructures.
A clarification of this point would be very useful.

Note that $\Eapp(\epsmb)$ is of special importance since it bridges the separation between
dynamics (diffusion constant $D(T)$) and thermodynamics (population of $\epsmb$).
Clearly, an understanding of $\Eapp(\epsmb)$ from basic principles is highly desirable.
It is plausible that the simple form $\Eapp(\epsmb)=\epsth-\epsmb$
can only be expected for a system acting as a completely correlated entity.
In contrast, two independently relaxing subsystems should
generally show a weaker dependence of $\Eapp(\epsmb)$ on $\epsmb$.
This can be seen by a very simple argument.
\newcommand{\epsmbone}{\epsmb^\ind{(1)}}
\newcommand{\epsmbtwo}{\epsmb^\ind{(2)}}
Consider two independent, identical systems, with MB energies $\epsmbone,\epsmbtwo$
and activation energy $\tilde\Eapp(\epsmb^\ind{(1,2)})$.
What can be said about $\Eapp(\epsmb)$ of the union of these systems, at MB energy
$\epsmb=\epsmbone+\epsmbtwo$?
In the limit of low temperatures, the apparent activation energy
is given by $\min(\tilde\Eapp(\epsmbone),\tilde\Eapp(\epsmbtwo))$.
A proper average over different realizations $\epsmbone$, $\epsmbtwo=\epsmb-\epsmbone$
yields $\Eapp(\epsmb)$ of the combined system.
Instead of carrying out this average, we use the fact that $\tilde\Eapp(\epsmb)$ is a monotonous
function and estimate
\EQU{0\le\min(\tilde\Eapp(\epsmbone),\tilde\Eapp(\epsmbtwo))\le\tilde\Eapp(\epsmb/2).}{}
Thus, $0\le\Eapp(\epsmb)\le\tilde\Eapp(\epsmb/2)$, which means that
the combined system shows a weaker dependence on $\epsmb$ than a single copy.
For a reasonable PEL topology, one would
expect $|\d\Eapp(\epsmb)/\d\epsmb|\le1$, because
barriers should not mount up more than one descents in the PEL.
Since the $\epsmb$-dependence of $\Eapp$ becomes weaker for
larger systems, it in turn should increases towards smaller~$N$.
As a speculation, this might open a way of estimating the size of cooperative regions.

The results shown in \figref{FIGDIFFTAU}, obtained via \equtwo{EQUDIFFTDPEL}{EQUTAUEPSMB},
demonstrate the use of the present work. From PEL barriers ($\Eappest(\epsmb)$)
and thermodynamics ($p(\epsmb;T)$) we are able to produce a reasonable estimate
of dynamics.
An overall proportionality factor $1/\tau_0$ remains as an adjustable parameter,
since it could not be predicted from PEL structure.
As discussed in section~\ref{SECMTAU}, one may use $p(\eps;T)$ instead of $p(\epsmb;T)$,
since they are nearly identical. This is very convenient,
because upon contructing $p(\eps;T)$,
no information about dynamics is needed.
The breakdown of the Arrhenius form of $\tauepsmbT$
above $2T_c$ limits our description to the temperatures $T\le2T_c$.
In any event, we would not have dared to make quantitative statements on the
basis of the hopping picture above the landscape-influenced
temperature regime. At $T=1$, for instance, we have $\eww{\epsmb}_T\approx-289$
(\figref{FIGEPS}), where we already find PEL barriers $\Eapp(-289)\approx1=\kB T$
(\figref{FIGTAU}).

From the fact that we could quantitatively relate MB lifetimes to PEL barriers below $2T_c$
and the results from section~\ref{SECACT},
we see that there exist activated barrier crossing events significantly above $T_c$.
As shown before~\cite{Doliwa:341}, these escape processes from stable
MBs determine the temperature dependence of the diffusion constant
also above $T_c$.
Thus, the general statement that hopping events are not relevant there
(see, e.g.~\cite{Franosch:385}) is not correct for the BMLJ system.
This implies that the ideal MCT can be
applied to systems for which activated processes {\it determine}
the time-scale of relaxation. Thus it seems that the theoretical
description of the cage effect in terms of structural quantities, as
done in MCT, works independent of whether the cage effect is purely
entropic (like in hard-sphere systems) or is to a large degree based on
barrier-crossing events.

Moreover, with the help of the unbiased quantity $\pbb(T)$, we were able to
measure the population of basin borders.
No indication for an abrupt change of relaxation mechanism could be observed in $\pbb(T)$;
in contrast, the separation of intra- and inter-basin motion
seems to happen rather smoothly (see \figref{FIGPBB}).
Thus, there is no qualitative change around $T_c$.

We finally discuss the relation of our work to
the instantenous normal mode approach (INM)
which considers the number of 'diffusive modes', $f_\ind{diff}$, 
to be at the physical basis of diffusion~\cite{Scala:375,Scala:376,LaNave:265}.
From the directions corresponding to negative eigenvalues of the Hessian
$H(x(t))$ (unstable directions), one filters out the 'diffusive' directions.
Considering the energy profile on the straight lines along the unstable directions,
La~Nave~et~al. observed extremely small barriers, indicating
completely 'entropic' dynamics at the considered temperatures~\cite{Scala:376}.
This conclusion, though reached for a model of supercooled water, is in contrast to
our findings of the relevance of energetic barriers.
A possible key to this apparent contradiction is that
$f_\ind{diff}$ is directly related to the fraction of time spent in 'mobile'
regions of configuration space.
In contrast, we have concentrated on the durations of the
stable, immobile structures. As the consequence of longer and longer residences in deep
MBs, the mobile fraction becomes smaller and smaller.
Thus, one observes a relation between $D(T)$ and $f_\ind{diff}$,
although it is the long trapping times which are the reason for the slowing down of dynamics.

We further note that the MB concept is not implemented in the INM approach.
Supercooled water, e.g., exibits very pronounced MB correlations in
the time series of minima, even for a 'large' system of 216 particles~\cite{Giovambattista:288}.
Generally, fragile glass formers are expected to have a 'rugged' PEL,
i.e. exibit extensive superstructures of minima~\cite{Stillinger:333}.
In view of this insight, the success of INM analyses for the latter type of systems
is quite surprising.


\section{Conclusion.}
\label{SECCONCLUSION}

Our goal in this paper was to shed some light on the temperature dependence
of the diffusion constant.
In our previous work~\cite{Doliwa:341}, metabasins turned out as a useful
concept that reduces correlations of subsequent PEL-hopping events.
Taking seriously these correlations, the present investigation
went a step further into this direction, by relating the temperature
dependence of relaxation to the depths of these multi-minima superstructures.
We have shown in this paper that a quantitative link between
PEL structure and dynamics is possible above $T_c$.
However, our approach is still phenomenological at this stage:
we are far from predicting $\Eapp(\epsmb)$
from more general PEL properties or even the interaction potentials.
To achieve this is a major challenge, implying a deep understanding of energy landscape
topology.

Further insights might be obtained by unveiling the {\it real-space} aspects
of MB relaxation. Here, the correspondance of MBs to the cage effect should serve
as the guiding principle. 
An interesting question along this line would be if some of the real-space phenomena
found in supercooled liquids (e.g. the string-like motion discovered by Donati~et~al.~\cite{Donati:98})
can be traced back to energy landscape features.

We are pleased to thank M.~Fuchs, D.R.~Reichman, and H.W.~Spiess for helpful discussions.
This work has been supported by the DFG, Sonderforschungsbereich 262.

\bibliographystyle{apsrev}
\bibliography{act}

\end{document}